\documentclass[draft]{agujournal2019}
\pdfoutput=1
\usepackage{url} 
\usepackage{lineno}
\usepackage{soul}
\usepackage{float}

\usepackage{amsmath} 
\usepackage{amssymb}
\draftfalse

\journalname{JGR: Space Physics}

\begin{document}

%
%


\title{
New Models of Jupiter's Magnetopause and Bow Shock through the \textit{Juno} Prime Mission: Probabilistic Location, Shape, and Internally-driven Variation}

%
%




\authors{M. J. Rutala\affil{1}, C. M. Jackman\affil{1}, C. K. Louis\affil{2}, A. R. Azari\affil{3, 4, 5}, F. Bagenal\affil{6}, S. P. Joy\affil{7}, W. S. Kurth\affil{8}, T. B. Keebler\affil{9}, R. S. Giles\affil{10}, R. W. Ebert\affil{10, 11}, C. F. Bowers\affil{1}, M. F. Vogt\affil{12}}

\affiliation{1}{Astronomy \& Astrophysics Section, School of Cosmic Physics, Dublin Institute for Advanced Studies, DIAS Dunsink Observatory, Dublin D15 XR2R, Ireland}
\affiliation{2}{LIRA, Observatoire de Paris, Université PSL, CNRS, Sorbonne Université, Université de Paris, Meudon, France}
\affiliation{3}{Department of Physics, University of Alberta, Edmonton, AB, Canada}
\affiliation{4}{Department of Electrical and Computer Engineering, University of Alberta, Edmonton, AB, Canada}
\affiliation{5}{Alberta Machine Intelligence Institute, Edmonton, AB, Canada}
\affiliation{6}{Laboratory for Atmospheric and Space Physics, University of Colorado Boulder, Boulder, CO, USA}
\affiliation{7}{Department of Earth, Planetary, and Space Science, University of California, Los Angeles, CA, USA}
\affiliation{8}{Department of Physics and Astronomy, University of Iowa, Iowa City, IA, USA}
\affiliation{9}{Department of Climate and Space Sciences and Engineering, University of Michigan, Ann Arbor, MI, USA}
\affiliation{10}{Space Science Division, Southwest Research Institute, San Antonio, Texas, USA}
\affiliation{11}{University of Texas at San Antonio, San Antonio, TX, USA}
\affiliation{12}{Planetary Science Institute, Tucson, AZ, USA}




\correspondingauthor{Matthew J. Rutala}{mrutala@cp.dias.ie}




\begin{keypoints}
\item New, data-driven boundary models for Jupiter's magnetopause and bow shock surfaces are derived using Bayesian statistical techniques.
\item The estimated magnetopause is polar-flattened (${\sim}13\%$) and asymmetric in the dawn-dusk direction; the estimated bow shock is mostly symmetric.
\item \textit{Juno} will likely spend $10\times$ longer in the solar wind during the extended mission compared to estimates from earlier boundary models.
\end{keypoints}

%
%

%
%


\begin{abstract}

The interaction between Jupiter's magnetosphere and the solar wind is not well-constrained: while internal energetic plasma processes are thought to dominate plasma circulation, the solar wind nonetheless exerts significant control over the shape and scale of the whole structure. To better constrain this interaction, we derive new functional forms for Jupiter's magnetopause and bow shock using data from the \textit{Ulysses}, \textit{Galileo}, \textit{Cassini}, and \textit{Juno} missions and calibrated solar wind estimates from the Multi-Model Ensemble System for the Heliosphere (MMESH). We design an empirical Bayesian model to estimate the locations of the boundaries using a Markov-chain Monte Carlo (MCMC) algorithm, expanding our model to sample all times, not only boundary crossing events. The boundary surfaces which best describe the data are thus estimated without the need for a full, physics-based magnetohydrodynamic (MHD) treatment of the Jovian magnetosphere and the additional assumptions required for such. The new magnetopause model exhibits significant polar flattening and dawn-dusk asymmetry, and includes a narrowing of the magnetotail when compared to previous models. The new bow shock model is largely axisymmetric. Both boundary models describe surfaces which lie closer to Jupiter than previous models, which has important implications for the modern picture of Jupiter's dynamic magnetosphere and the expected science results of current and upcoming Jupiter-bound spacecraft. Applying these models to \textit{Juno}'s trajectory, we estimate that the spacecraft should be expected to spend ${\sim}19\%$ of each orbit in the magnetosheath and ${\sim}4\%$ of each orbit in the solar wind starting from Perijove 64 (PJ64, 21 July 2021).

\end{abstract}

\section*{Plain Language Summary}
The solar wind, a perpetual stream of plasma and magnetic field lines flowing outwards from the Sun in all directions, interacts with all of the planets as it flows past them. Planets with their own internal magnetic fields create bubbles within this solar wind flow---magnetospheres---which are supported by the balance of pressures inside and outside. At most planets, a second boundary forms around the magnetosphere---a bow shock---caused by the fast-flowing solar wind suddenly stopping as it hits the magnetosphere. The shapes and sizes of a planet's magnetosphere and bow shock are continuously changing as the solar wind changes, and understanding these changes is important for determining how mass and energy move through space near that planet. Here, we investigate the size and shape of Jupiter's magnetosphere and bow shock using spacecraft measurements and solar wind models, and define new functions that describe how the shapes change with the solar wind. The new shape of the magnetosphere is asymmetric, being flatter near the poles and inflated near dusk. The sizes of both the magnetosphere and the bow shock are smaller than previously thought, meaning that current missions to Jupiter, like \textit{Juno}, will leave the magnetosphere and bow shock more frequently.

%
%
\section{Introduction}

The Jovian system has a complex interaction with the solar wind. Fundamentally, Jupiter's magnetosphere is supported by internal pressures which balance the external pressure, which is itself ultimately derived from the solar wind. The isobaric surface between these two regions defines the magnetopause. The solar wind is deflected and decelerated around this boundary, increasing in temperature and density; as this solar wind is both super-Alfvénic (Alfvén Mach number $M_{A} \gg 1$) and super-fast (fast-mode magnetosonic Mach number $M_{MS} \gg 1$) \mbox{\cite{Ebert2014}}, the transition from undisturbed solar wind to deflected plasma is discontinuous and another boundary, the bow shock, forms.

The shapes and scales of both of these boundaries are controlled by the interplay between the pressures external and internal to the magnetosphere. Externally, the pressure is dominated by the solar wind dynamic pressure $p_{SW}$, which has been measured to vary over more than two orders of magnitude near Jupiter \cite{Jackman2011, Ebert2014}. Internally, the pressure cannot be described by a single component; at Jupiter, the dynamic, magnetic, and thermal plasma pressures are all important, and vary in relative importance with location and time \cite<e.g.>{Hill1974, Connerney1981_model, Lepping1980, Kivelson2005, Huscher2021, Wang2024_results}. 

Modeling the shape and scale of these boundaries is a challenging problem, and is made yet more difficult by the rapid changes these boundaries undergo and the relative paucity of boundary-crossing events by spacecraft in the expansive Jovian system \cite<e.g.>{Ness1979a, Ness1979b, Lepping1980, Slavin1985, Huddleston1998b, Huddleston1998a, Joy2002}. Statistically investigating the possible shapes, scales, and behaviors of these boundaries in response to the solar wind is not only a practical approach to this challenge, but also yields deep insights into Jupiter's magnetosphere and the processes embedded within. The degree to which the solar wind is responsible for moving plasma and energy throughout the magnetosphere is fundamental to our understanding of Jupiter's near-space environment, yet remains contested \cite<e.g.>{McComas2007, Cowley2008_comment, Delamere2010}. Solar wind coupling with the Jovian magnetosphere may influence the planet directly, for instance by modulating Jupiter's powerful ultraviolet (UV) auroral storms \cite{Bonfond2020, Bonfond2021, Yao2022, Yao2024}; the nature of this interaction remains an open debate \cite<e.g.>{Kotsiaros2019, Lorch2020, Nichols2020, Nichols2022, AlSaati2022, Kamran2022, Rutala2024}, in part due to the difficulty in constraining the degree of solar wind influence over Jupiter's magnetosphere.

Here, we address this demanding problem by adopting an empirical Bayesian model and employing a Markov-chain Monte Carlo (MCMC) algorithm to derive new boundary models that fully leverage the available spacecraft data at all times, not just during boundary crossing events, following \citeA{Joy2002}. The approach here differs from earlier work, as will be discussed in Section \ref{sec: Background}, in two key ways: by incorporating the wealth of new spacecraft data available since the most recent boundary models were developed; and by being independent of complex, magnetohydrodynamic (MHD) models of Jupiter's magnetosphere and its interaction with the solar wind. The assorted spacecraft measurements, solar wind propagation models, and mathematical forms for the boundary shapes will be introduced in Section \ref{sec: Data and Models} prior to discussing how these data and models are used to derive boundary forms in Section \ref{sec: Analysis}. Finally, the new boundary models derived here will be presented in Section \ref{sec: Results}, where they will be interpreted relative to the crossing data and previous boundary models.

\section{Prior Exploration of Jupiter's Magnetospheric Boundaries}\label{sec: Background}
Jupiter's magnetosphere is uniquely large, rapidly-rotating, and plasma-dense compared to other magnetized solar system objects. It possesses an internal source of dense plasma, ultimately stemming from the volcanic moon Io, coupled with a rapid rate of planetary rotation, giving rise to relatively large, internal thermal plasma and dynamic pressures. Near the planet, this dense thermal plasma is accelerated in the direction of planetary rotation by magnetosphere-ionosphere coupling currents \cite{Hill1979, Hill2001, Cowley2001, Southwood2001, Ray2010}, driven along magnetic field lines toward a balance point $9.5^\circ$ off the rotational equator toward the magnetic equator by centrifugal forces \cite{Hill1974, Bagenal1981, Phipps2021}, and propelled radially outward by flux-tube interchange motion \cite{Siscoe1981, Vasyliunas1983, Kivelson2005, Dougherty2017}. These factors yield a dynamic plasma environment with a geometry which is significantly extended in the equatorial plane. In the middle magnetosphere ($10-40$ $\textrm{R}_{\textrm{J}}$), magnetodisk densities and temperatures have recently been observed with the \textit{Juno} spacecraft to vary by a factor of $100$ and velocities by a factor of $\sim3$ \cite<e.g.>{Wang2024_results}, with both large-scale, orbit-to-orbit and small-scale, intra-orbit variations measured \cite{Huscher2021}. Combined with the ever-changing solar wind, these internal dynamics---which may themselves be influenced by the solar wind---result in the variable locations and shapes of Jupiter's magnetospheric boundaries.

The variability of Jupiter's magnetopause was first documented during the Pioneer 11 flyby, when the spacecraft crossed the subsolar magnetopause at ${\sim}100 \ \mathrm{R}_\mathrm{J}$ and again at ${\sim}50 \ \mathrm{R}_\mathrm{J}$ on four occasions. Three of these crossings coincided with elevated solar wind flow speed and pressure as propagated from Pioneer 10, offset from Pioneer 11 by $12.9^\circ$ heliolongitude and $1.4^\circ$ heliolatitude \cite{Smith1978, Smith1981}. While slightly lower variability in magnetopause crossing distance was observed during the Voyager flybys \cite{Lepping1980}, it is evident that Jupiter's magnetopause is nonetheless highly compressible: the boundary shrinks $\gtrsim5\times$ more easily than the Earth's under increased solar wind pressure \cite{Smith1978}, and can additionally shrink due to strictly internal variation in the plasmadisk \cite{Smith1978, Engle1980}. 

At low latitudes, the plasmadisk provides the balancing pressure inside the magnetopause, while near the poles, the boundary is expected to be flattened compared to the dawn flank \cite<e.g.>{Hill1974}. Taking $r_{dawn}$ to be the distance to the dawn boundary (i.e., the boundary location as measured in the equatorial plane at 0600 hours local solar time) and $r_{polar}$ to be the distance to the polar boundary (i.e., the boundary location as measured in the direction aligned with Jupiter's spin normal), the polar magnetopause is expected to be flattened by a factor $f_{polar} = (r_{dawn} - r_{polar})/r_{dawn} = {\sim}20\%-40\%$ \cite{Engle1980, Connerney1981_model, Lepping1981, Ranquist2020}. As the thickness of the plasmadisk varies from dawn to dusk, an asymmetry along this axis is also expected; the nature of this asymmetry depends on the exact plasma parameters in both regions \cite{Lepping1980, Kivelson2002, Kivelson2005}. In the antisunward direction, the Jovian magnetotail has been observed to reach extremely large scales, spanning ${\gtrsim}300 \ \mathrm{R}_\mathrm{J}$ along the dawn-dusk axis \cite{Ness1979b, McNutt2007, Ebert2010} and $\gtrsim9000 \ \mathrm{R}_\mathrm{J}$ ($\gtrsim4.3$ AU) in the antisolar direction \cite{Kurth1982, Lepping1982, Lepping1983, McComas2007_NH}. The variability of the Jovian bow shock, forming in response to the magnetopause as an obstacle to the flow of the solar wind, is directly tied to the form and compression state of the magnetopause boundary; the bow shock is, however, likely to be more blunt than the comparatively streamlined shape of the underlying magnetopause \cite<e.g.>{Joy2002}.

The Jovian magnetopause and bow shock are thus constantly in motion as they react to both internal and external changes; averaging over time, however, a statistical view of the shape, scale, and behavior of these boundaries can be taken. Hyperbolic \cite{Ness1979a} or parabolic \cite{Ness1979b, Lepping1980} forms for the magnetopause, assumed to be axisymmetric about the Sun-Jupiter axis and static relative to the solar wind, were derived based on the flybys of Voyagers 1 and 2; for the same flybys, the bow shock was modeled as strictly hyperbolic and similiarly axisymmetrical and static \cite{Ness1979a, Ness1979b, Lepping1980}. In each of these forms, the subsolar standoff point of the boundary $r_{SS}$ is situated at the vertex of the curve without aberration, as the solar wind is primarily directed radially outward. By combining the Voyagers 1 and 2 crossing data with those of the earlier Pioneers 10 and 11, an elliptical form for the magnetopause and hyperbolic form for the bow shock were derived by \citeA{Slavin1985}. These forms are significant for not being presupposed; rather, an axisymmetric equation describing any conic section was fit to the crossing data, after correcting the data to an average solar wind dynamic pressure value. On the basis of these models, the subsolar standoff distance of either boundary $r_{SS}$ was found to relate to the solar wind dynamic pressure $p_{SW}$ as $r_{SS} \propto p_{SW}^{-1/4.4} - p_{SW}^{-1/4}$, an expected polar flattening of $f_{polar}{\approx}40\%$ was recovered, and the typical ratio of the bow shock and magnetopause subsolar standoff distances ($R_{SS, BS}$ and $R_{SS, MP}$, respectively) was found to be $r_{SS,BS}/R_{SS,MP}{\approx}1.25$ \cite{Slavin1985}. The same boundary model form, applied to early \textit{Galileo} crossing data, recovered comparable estimates for an elliptical magnetopause and a hyperbolic dayside bow shock: $r_{SS} \propto p_{SW}^{-1/4.5}$ and $r_{SS,BS}/r_{SS,MP}{=}1.12{\sim}1.2$; these \textit{Galileo}-era models, however, predict less flattening, with $f_{polar}=3\%{\sim}13\%$ \cite{Huddleston1998a, Huddleston1998b}.

Most recently, \citeA{Joy2002} utilized the boundary crossings from the \textit{Voyagers} 1 and 2, \textit{Ulysses}, and \textit{Galileo} spacecraft to derive boundary models based on a slightly different functional form, which is discussed in more detail in Section \ref{sec: Boundary Shape Models}. Rather than fitting to the location of the boundary crossings, this latest model was instead derived based on the fraction of time spent inside each boundary as a function of distance from Jupiter and employed a Bayesian statistical framework to interpret these data. The boundary shapes in the \citeA{Joy2002} model were then derived from three-dimensional (3D) MHD models of Jupiter's magnetosphere, with the scales of these boundaries determined from the statistical analysis of spacecraft data, with an assumed relationship between $p_{SW}$ and $R_{SS}$ of $R_{SS} \propto p_{SW}^{-1/4}$.This model recovers a polar flattening of $f_{polar} = 5\%{\sim}8\%$ and a standoff distance ratio between the bow shock and magnetopause of $R_{SS,BS}/R_{SS,MP}={\sim}1.2-1.3$. The standoff distance ratio in this model is modulated by a bimodally distributed magnetopause surface, wherein the magnetopause surface probability density is double-peaked as a function of distance from Jupiter. While more recent estimates suggest that this model likely underestimates the polar flattening \mbox{\cite{Ranquist2020}}, the bimodality of the Jovian magnetopause has been confirmed with a k-means analysis \cite{Collier2020}. Jupiter's bimodal magnetopause may be explained by the distribution of solar wind dynamic pressures near Jupiter \mbox{\cite{McComas2014}}. This bimodality may be common to the outer planets more generally, as the Kronian magnetopause has been found to have a similar distribution \cite{Achilleos2008}.

\section{Data and Models}\label{sec: Data and Models}

For our approach to modeling the shape, scale, and response of Jupiter's magnetospheric boundaries, we require a catalog of spacecraft boundary crossings and ephemerides (Section {\ref{sec: Spacecraft Measurements}}), a model of the solar wind (Section {\ref{sec: Solar Wind Model}}), and functional forms for the boundaries themselves (Section {\ref{sec: Boundary Shape Models}}); these three components will be discussed in more detail here.

\subsection{Spacecraft Measurements}\label{sec: Spacecraft Measurements}

To date, nine spacecraft have transited through Jupiter's magnetopause and bow shock: \textit{Pioneers} 11 and 12, \textit{Voyagers} 1 and 2, \textit{Ulysses}, \textit{Galileo}, \textit{Cassini}, \textit{New Horizons}, and \textit{Juno}. Here, we have used data from the \textit{Ulysses}, \textit{Galileo}, and \textit{Cassini} spacecraft, and from the \textit{Juno} spacecraft through the end of its prime mission (through 30 July 2021) \cite{Bame1992, Kurth2002, Achilleos2004, Svenes2004, Louis2023}. The earlier flybys of the \textit{Pioneers} and \textit{Voyagers} are not included due to the incompleteness of near-Earth and solar observations from which to model the solar wind, which will be required for the Bayesian estumation method outlined in Section \ref{sec: Analysis}. The \textit{New Horizons} flyby is not used here as the boundary crossings it recorded are too scarce to significantly constrain either magnetospheric boundary in the region of interest here ($r \lesssim 500 \ \mathrm{R}_\mathrm{J}$). The trajectory of \textit{New Horizons} took the spacecraft deep into Jupiter's magnetosphere and through the magnetotail, where it only measured the bow shock and magnetopause once within our region of interest---though, when \textit{New Horizons} left the magnetosphere through the distant magnetotail ($>2000 \ \mathrm{R}_\mathrm{J}$), it encountered the magnetopause several times \cite{McNutt2007}.

\begin{table}[!htb]
    \begin{tabular}{llccc}
    \hline
    & Spacecraft & First Encounter & Final Encounter & Encounters \\
    &            & (Year-DoY HH:MM) & (Year-DoY HH:MM) &           \\
    \hline
    Bow Shock & \textit{Ulysses}$^{a}$ & 1992-033 17:33 & 1992-047 07:55 & 4 \\                
              & \textit{Galileo}$^{b*}$ & 1996-148 10:17 & 2002-321 12:00 & 120 \\            
              & \textit{Cassini}$^{c}$ & 2000-363 04:19 & 2001-058 10:45 & 30 \\             
              & \textit{Juno}$^{e,f}$ & 2016-176 08:16 & 2017-012 14:05 & 75 \\       
    \hline
    \multicolumn{4}{r}{\textit{Total}:} & 229 \\
    \hline
    Magnetopause & \textit{Ulysses}$^{a}$ & 1992-033 23:00 & 1992-045 20:45 & 12 \\
                 & \textit{Galileo}$^{b*}$ & 1996-145 06:25 & 2002-316 10:13 & 142 \\
                 & \textit{Cassini}$^{b,d}$ & 2001-009 12:50 & 2001-010 20:35 & 4 \\
                 & \textit{Juno}$^{e,f}$ & 2016-177 21:20 & 2018-216 08:51 & 215 \\
    \hline
    \multicolumn{4}{r}{\textit{Total}:} & 369 \\
    \hline
    \multicolumn{5}{l}{$^a$ from \citeA{Bame1992}} \\
    \multicolumn{5}{l}{$^b$ from \citeA{Kurth2002}} \\
    \multicolumn{5}{l}{$^c$ from \citeA{Achilleos2004}} \\
    \multicolumn{5}{l}{$^d$ from \citeA{Svenes2004}} \\
    \multicolumn{5}{l}{$^e$ from \citeA{Hospodarsky2017}} \\
    \multicolumn{5}{l}{$^f$ from \citeA{Louis2023}} \\
    \multicolumn{5}{l}{$^*$ from \citeA{Joy2010}; reproduced in S1.} \\
    
    \end{tabular}
\caption{Summary of spacecraft encounters with Jupiter's bow shock and magnetopause}
\label{tab: Crossings Summary}
\end{table}

\begin{figure}[!th]
    \includegraphics[width=0.99\textwidth]{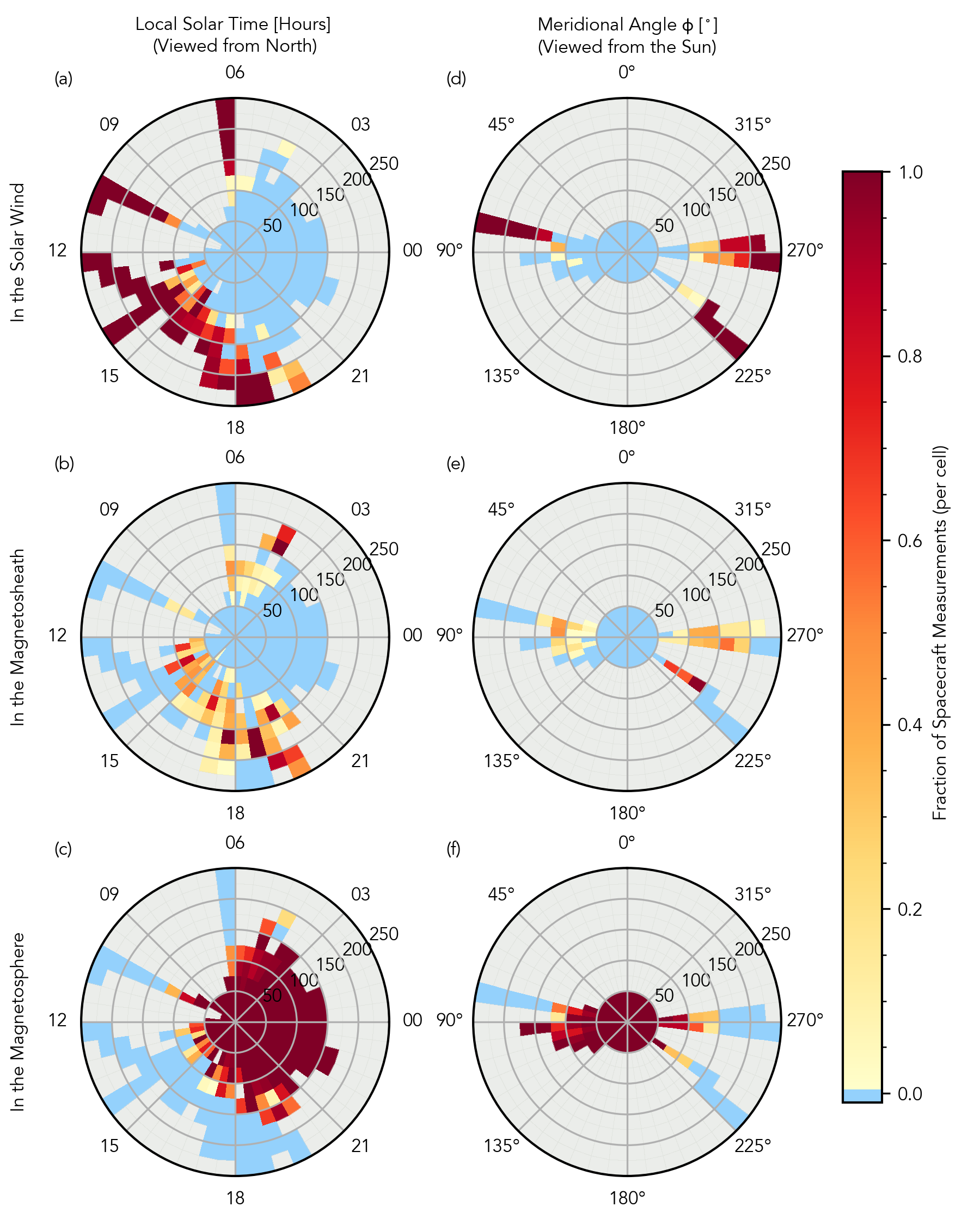}
    \caption{Residence plots showing the fraction of time spent in each magnetospheric region by all spacecraft, combined. The first column (panels a-c) shows this fraction as a function of local solar time (LST) and includes all data within $\pm50 \ \mathrm{R}_\mathrm{J}$ of the $z_{JSS}{=}0$ (equatorial) plane; the second column (panels d-f) shows this fraction as a function of meridional angle (positive counterclockwise from North toward dawn in the $x_{JSS}{=}0$ plane, as viewed from the Sun) and includes all data within $\pm50 \ \mathrm{R}_\mathrm{J}$ of the $x_{JSS}{=}0$ plane. Grid cells are colored based on the fraction of time spent in the specified region relative to the total time spent within the specific cell, following the colorbar to the right; cells where no spacecraft measurements have been recorded (gray) and where a certain region has never been observed (blue) are colored separately.}
    \label{fig:Residence}
\end{figure}

All crossing lists employed here identified boundary crossings from discontinuities in the magnetic field, with the lists of \citeA{Kurth2002}, \citeA{Svenes2004}, \citeA{Joy2010}, \citeA{Hospodarsky2017}, and \citeA{Louis2023} additionally confirming the boundary crossing events with plasma data when available. Within each flyby or sufficiently high orbital pass, spacecraft typically observe each boundary multiple times, as the boundaries fluctuate over small spatial scales (${\sim}10 \ \mathrm{R}_\mathrm{J}$) much more rapidly than typical spacecraft speeds of $1\sim6$ km/s \cite<e.g.>{Zhang2018, Ma2022}. Table \ref{tab: Crossings Summary} summarizes the times of the first and final boundary encounters of these spacecraft with both the bow shock and the magnetopause (in year--day of year (DoY) format), as well as the total number of encounters. The time spent by all spacecraft in each of the regions bounded by the bow shock and magnetopause-- the solar wind, the magnetosheath, and the magnetosphere-- is illustrated by the residence plots in Figure \ref{fig:Residence}. The combined crossings of \textit{Ulysses}, \textit{Galileo}, \textit{Cassini}, and \textit{Juno} used here have representative coverage in local solar time (LST), as illustrated by Figure \ref{fig:Residence}a-c, but have limited coverage of the polar extent of the system for $r > 40 \ \mathrm{R}_\mathrm{J}$, as illustrated by Figure \ref{fig:Residence}d-f.

\subsection{Solar Wind Model}\label{sec: Solar Wind Model}
To determine the relationships between the boundary shapes, scale, and behavior with the solar wind, the previously discussed boundary crossing measurements must then be combined with one of two broad classes of model: either a 3D model of the magnetosphere under different solar wind conditions, as employed previously by \mbox{\citeA{Joy2002}}, or a model of the time-varying solar wind. Here, the second method will be used, with the most probable solar wind conditions provided by the Multi-Model Ensemble System for the outer Heliosphere (MMESH, \mbox{\citeA{Rutala2024_MMESH}}). MMESH is a meta-modeling framework within which the timing uncertainties of physical solar wind models are first calculated by comparison to in-situ solar wind data, then modeled as a response to physical parameters via multiple linear regression. Modeling the timing uncertainties as such allows the uncertainties to be estimated when in-situ solar wind data are unavailable. The resultant models are then shifted to account for offsets in predicted arrival time, with propagated uncertainties at each time step estimated in a consistent manner across models. These final models are then combined to create a single multi-model ensemble (MME) with uncertainties, which tends to outperform any individual input model when compared to data. Opting to rely on this model of the time-varying solar wind near Jupiter is preferred over a 3D MHD simulation of Jupiter's magnetosphere to limit the number of free parameters entering into the modeled boundary forms.

The MMESH framework has been updated for better performance and to allow direct sampling of the predicted distributions of solar wind parameters; these updates, and the performance of the resulting MME, are discussed in detail in \ref{appendix: MMESH}. Following these updates, we created a solar wind MME based on the HUXt \cite{Owens2020, Barnard2022}, MSWIM2D \cite{Keebler2022}, and Tao+ \cite{Tao2005} solar wind propagation models and using in-situ solar wind measurements from the \textit{Ulysses} and \textit{Juno} spacecraft, contemporaneous with the propagation models, for comparison. The measured timing uncertainties were then regressed to the smoothed F10.7 radio flux (a proxy for the solar cycle), the solar recurrence index, the difference in heliolongitude and heliolatitude between the model target and the Earth, and the modeled solar wind flow speed. This MME covers the same spans as listed in Table \ref{tab: Crossings Summary}, and thus can be used to estimate the solar wind dynamic pressure $p_{SW}$ during any of these spacecraft boundary encounters. The overall performance of the MME is summarized in Figure \ref{fig:MMEPerformance}; most importantly, the MME captures the mean and variance of measured $p_{SW}$ best (e.g., Figure \ref{fig:MMEPerformance}e) and has the highest correlation coefficient with measured $p_{SW}$ (e.g., Figure \ref{fig:MMEPerformance}f) when compared against the other solar wind models tested.

\subsection{Boundary Shape Models}\label{sec: Boundary Shape Models}
\begin{figure}[!th]
    \centering
    \includegraphics[width=0.5\textwidth]{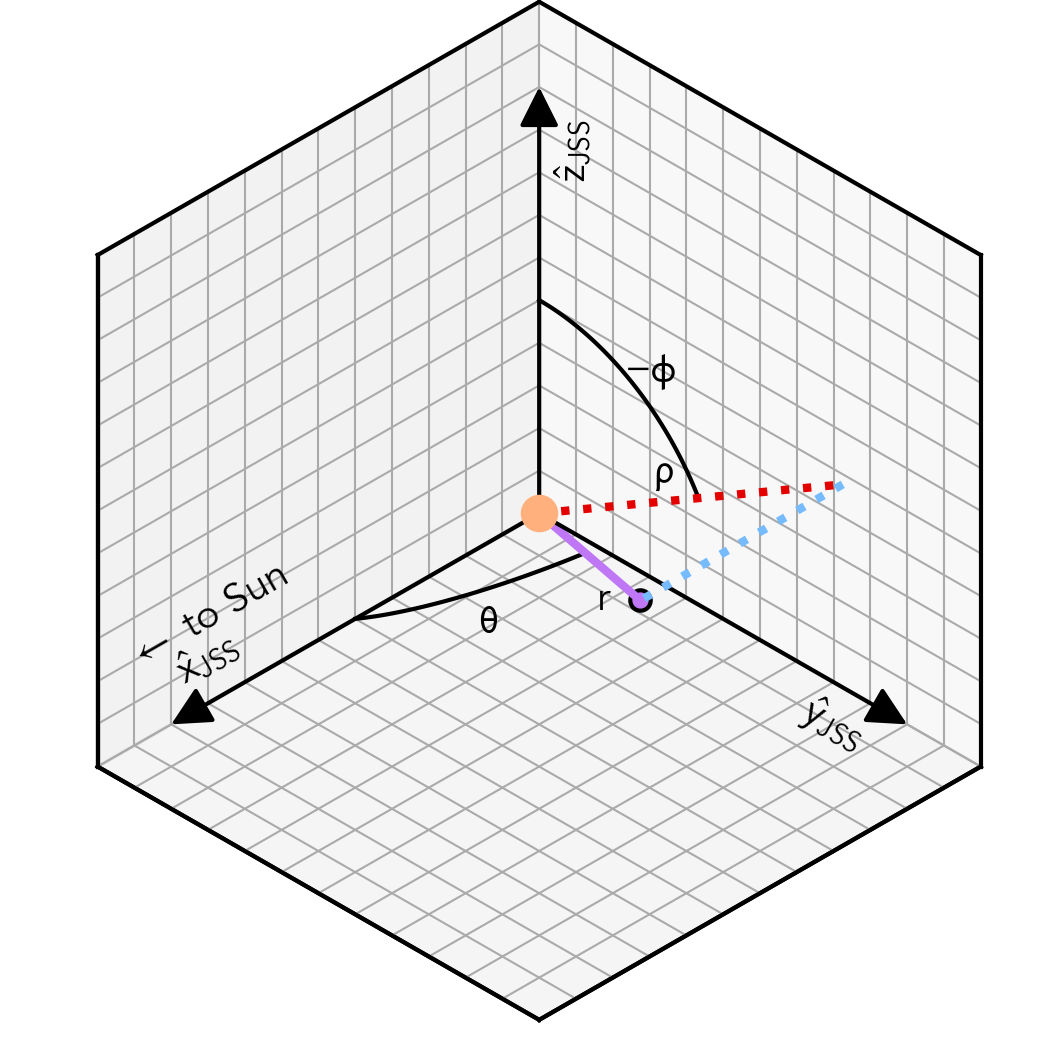}
    \caption{Schematic representation of the sun-oriented ``spherical-solar'' coordinate system implemented here relative to the Jupiter-de-Spun-Sun (JSS) coordinate system. Points in this spherical-solar coordinate system are defined by the coordinates $(r, \theta, \phi)$: $r$ is the distance from the center of Jupiter ($(x_{JSS}, y_{JSS}, z_{JSS}) = (0, 0, 0)$); $\theta$ is an angle measured from $+\mathbf{x_{JSS}}$ (i.e., roughly from the sub-solar point) back toward the tail; and $\phi$ is an angle measured anticlockwise from North, from the perspective of the Sun (i.e., from $+\mathbf{z_{JSS}}$, increasing through $-\mathbf{y_{JSS}}$). This coordinate system conveniently exploits the expected zeroth-order symmetries of the magnetospheric boundaries.}
    \label{fig:Coords}
\end{figure}

The models of the magnetospheric boundary shapes and response to solar wind parameters used here are defined in either the Jupiter-de-Spun-Sun (JSS) coordinate system or the ``spherical-solar'' coordinate system based on those used for previous models \cite<e.g.>{Slavin1985, Shue1997, Huddleston1998b}. The JSS coordinate system is defined such that $+\mathbf{z_{JSS}}$ is aligned with the Jupiter spin vector, and, for a unit vector pointing from Jupiter toward the Sun $\mathbf{r_{Sun}}$, $+\mathbf{y_{JSS}} = \mathbf{r_{Sun}} \times +\mathbf{z_{JSS}}$ and $+\mathbf{x_{JSS}} = +\mathbf{y_{JSS}} \times +\mathbf{z_{JSS}}$. The spherical-solar coordinate system is illustrated in Figure \ref{fig:Coords}, and is defined by coordinates $r$, $\theta$, and $\phi$. The polar axis points along $+\mathbf{x_{JSS}}$ with polar angle $\theta$ measured from $+\mathbf{x_{JSS}}$, through the $y-z$ plane, toward $-\mathbf{x_{JSS}}$. The azimuthal angle $\phi$ is measured from North ($+\mathbf{z_{JSS}}$) positive through dawn ($-\mathbf{y_{JSS}}$), proceeding counterclockwise when viewing Jupiter from the Sun. This coordinate system leverages the expected zeroth-order symmetry of the system.

We compare three shape models to describe both the magnetopause and the bow shock. The first is the functional form of \citeA{Shue1997}, which was originally developed to describe the Earth's magnetopause \cite<e.g.>{Shue1998, Shue2002} but has since been applied to both the magnetopause and bow shock at other planets, including those of Mercury \cite{Winslow2013} and Saturn \cite{Kanani2010, Arridge2006}; we will refer to this model hereafter as ``S97*'', to make clear that we are estimating new parameters rather than using those described by \mbox{\citeA{Shue1997}}. For the state vector $\mathbf{x} = (\theta_{SC}, \phi_{SC}, p_{SW})$, describing the positions and pressure for which the distance to the boundary, $r_{b}$, is to be calculated, and model coefficient vector $\mathbf{c_{\mathrm{S97*}}} = (r_0, r_1, \alpha_0, \alpha_1)$, the S97* model has the form
\begin{equation}
\begin{split}
    r_{b} &= \mathcal{F}_{\mathrm{S97*}}(\mathbf{c_{\mathrm{S97*}}}, \mathbf{x}) = r_{SS} \left(\frac{2}{1 + \textrm{cos}(\theta)}\right)^{\alpha_f} \\
    r_{SS} &= r_0 \ p_{SW}^{r_1} \\
    \alpha_f &= \alpha_0 + \alpha_1 \ p_{SW} \\
\end{split}
\label{eqn:S97}
\end{equation}
in spherical-solar coordinates, where $r_{b}$ represents the radial distance to the boundary, $r_{SS}$ represents the sub-solar standoff distance of the boundary, and $\alpha_f$ represents the flaring of the tail. The S97* form is fundamentally an axisymmetric ellipsoid, with the addition of a flare parameter $\alpha_f$ which is absent in existing models of Jupiter's magnetospheric boundaries \cite<e.g.>{Slavin1985, Huddleston1998a, Huddleston1998b, Joy2002}. The flare angle $\alpha_f$ allows for the boundary to be open and increasingly wide, like a parabola or hyperbola ($\alpha_f > 0.5$), open and approaching a constant width asymptotically ($\alpha_f = 0.5$), or closed and of variable downtail length ($\alpha_f < 0.5$). A significant advantage of the S97* form, and magnetospheric boundary shapes derived from it, is the ease of interpreting the fitting parameters composing $\mathbf{c_{\mathrm{S97*}}}$: $r_0$ represents the subsolar standoff distance of the boundary for $p_{SW} = 1$ nPa and $r_1$ represents the sensitivity of the standoff distance to $p_{SW}$; $\alpha_0$ and $\alpha_1$ quantify the overall shape of the boundary surface and the degree to which it changes under changing $p_{SW}$. However, as the S97* form is axisymmetric, it is not expected to fit Jupiter's asymmetric magnetopause or bow shock particularly well; instead, a model with so few parameters provides a useful point of comparison to the more complex boundary models. Finally, we hold $r_1 = -0.25$ constant in exploring this model, given the abundance of evidence that the subsolar standoff distance is proportional to solar wind pressure as $r_{SS} \propto p_{SW}^{-0.25}$ \cite{Slavin1985, Huddleston1998a, Huddleston1998b, Joy2002}.

The second boundary form we will explore is a modified version of the S97* model, developed here and tailored to the known asymmetries of the Jovian system. For $\mathbf{x}$ and $\mathbf{c_{new}}=(r_0, r_1, r_2, r_3, \alpha_0, \alpha_1)$, the new model has the form 
\begin{equation}
    \begin{split}
        r_{new} &= \mathcal{F}_{new}(\mathbf{c_{new}}, \mathbf{x}) = r_{SS} \left(\frac{2}{1 + \textrm{cos}(\theta)}\right)^{\alpha_f} + r'_{b}\\
        r_{SS} &= r_0 \ p_{SW}^{r_1} \\
        \alpha_f &= \alpha_0 + \alpha_1 \ p_{SW} \\
        r'_{b} &= 
        \begin{cases}
        \sin^2(\theta/2) \left( r_2\sin^2(\phi)\right) \ p_{SW}^{r_1},& \text{if } 0 < \phi \leq \pi \\
        \sin^2(\theta/2) \left( r_3\sin^2(\phi)\right) \ p_{SW}^{r_1},& \text{if } \pi < \phi \leq 2\pi \\
        \end{cases}
    \end{split}
\label{eqn:new}
\end{equation}
in spherical-solar coordinates.
Equation \ref{eqn:new} differs from Equation \ref{eqn:S97} by the addition of a perturbation $r'_{b}$ to the distance to the boundary $r_b$ which allows for polar flattening and dawn-dusk asymmetry: $r_2$ and $r_3$ describe the inflation ($> 0$) or deflation ($< 0$) of the dusk and dawn flanks relative to the polar regions, respectively.
The shape of the dusk and dawn flanks of each boundary are explicitly independent in Equation \ref{eqn:new}. The remaining trigonometric terms ensure that the boundary is continuous in three dimensions and allow for direct interpretation of the effects of $r'_b$ on the overall boundary shape: along the dawn flank, $r'_b(\theta=90^\circ, \phi=90^\circ) = r_2/2$ and $r'_b(\theta=180^\circ, \phi=90^\circ) = r_2$; the same relationships hold for $r_3$ along the dusk flank $(\phi=270^\circ)$. The flaring parameters are assumed to be axisymmetric and, as with the S97 shape model, $r_1 = -0.25$ is held constant.

Finally, we will compare against the canonical \citeA{Joy2002} model, which we will refer to as the ``J02'' model to make clear that no changes from the original formulation have been made. The J02 model has the form
\begin{equation}
\begin{split}
    z_{b}^2 &= A + B \times x_{b} + C \times x_{b}^2 + D \times y_{b} + E \times y_{b}^2 + F \times x_{b} \times y_{b} \\
    A &= A_0 + A_1 p_{SW}^{-1/4} \\
    B &= B_0 + B_1 p_{SW}^{-1/4} \\
    C &= C_0 + C_1 p_{SW}^{-1/4} \\
    D &= D_0 + D_1 p_{SW} \\
    E &= E_0 + E_1 p_{SW} \\
    F &= F_0 + F_1 p_{SW} \\
\end{split}
\label{eqn:J02}
\end{equation}
in JSS coordinates, where $A_0, A_1, B_0, B_1, C_0, C_1, D_0, D_1, E_0,$ and $E_1$ represent the model coefficients. Alternative model parameters for the J02 model are not explored here due to difficulties constraining the model shape with the available data, as will be discussed in more detail after introducing the Bayesian inferencing formulation used to estimate model coefficients in Section \mbox{\ref{sec: Bayesian Fitting Methods}}.


\section{Analysis} \label{sec: Analysis}
\subsection{Data Post-Processing}\label{sec: Data Post-Processing}
With sufficiently many boundary crossing measurements to fully sample the boundary in different locations and under different conditions, the shape and scale of each boundary could be found from a simple regression, as has been done at the Earth \cite{Shue1997} and Mercury \cite{Winslow2013}, and as has been previously attempted at Jupiter \cite<e.g.>{Slavin1985, Huddleston1998a, Huddleston1998b}. Here, we have $369$ magnetopause crossing measurements and $229$ bow shock crossing measurements, as presented in Section \ref{sec: Spacecraft Measurements}. \citeA{Shue1997} use $533$ magnetopause crossings to derive the boundary form at Earth; \citeA{Winslow2013} use $1065$ magnetopause crossings and $1084$ bow shock crossings to derive both boundary forms for Mercury. Further, these models for the Earth and Mercury are both axisymmetric, effectively reducing the problem to a three-dimensional one ($r$, $\theta$, and $p_{SW}$); as Jupiter's boundaries are not expected to be axisymmetric, the added dimensionality of the problem at Jupiter (i.e., allowing the boundary models to vary in $\phi$) necessitates a different approach to successfully utilize the available crossing data.

Following \citeA{Joy2002}, we combine the crossing data with spacecraft trajectories to extract additional information about the location of each boundary $r_b$. The spacecraft boundary crossing events summarized previously are effectively instantaneous: during the event, the spacecraft intercepts the boundary and the position of the boundary surface is known exactly (i.e., $r_{SC} = r_b$). However, outside of these instants, the spacecraft trajectory can still provide information about the potential locations of the boundary. After a spacecraft has crossed a boundary in the inbound direction, we know that $r_{SC} < r_b < \infty$; similarly, after a spacecraft has crossed a boundary in the outbound direction, we know $0 < r_b < r_{SC}$. 

\begin{figure}[!ht]
    \includegraphics[width=\textwidth]{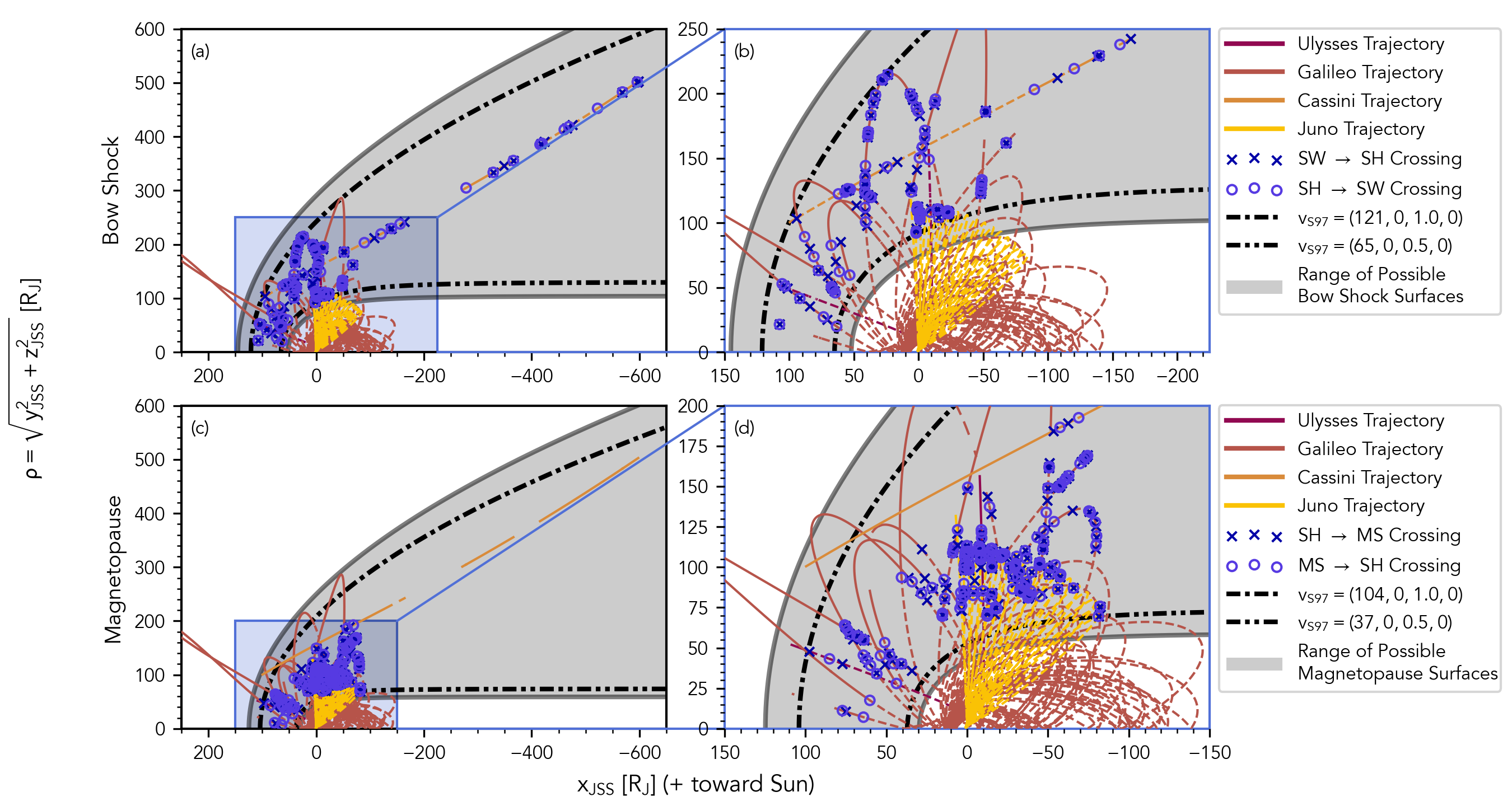}
    \caption{Orbital trajectories of the spacecraft used here (\textit{Ulysses}, \textit{Galileo}, \textit{Cassini}, and \textit{Juno}, as labeled) in the $x_{JSS}{-}\rho_{JSS}$ plane, with inward (blue ``x''s) and outward (purple ``o''s) boundary crossings marked; whether the spacecraft is outside (solid lines) or inside (dashed lines) the boundary, or if the location is not clear based on the crossing data (no line), is also demonstrated. Panels (a, c) include all bow shock and magnetopause crossings discussed in Section \ref{sec: Spacecraft Measurements} on the same scale, respectively. Panels (b, d) are the same, but re-scaled to make nearer trajectories and crossings easier to identify. Here, $\rho = \sqrt{y_{JSS}^2 + z_{JSS}^2}$ is the distance measured perpendicular from the $x_{JSS}$ axis, allowing the 3D trajectories and distribution of crossings to be represented in 2D. For each set of crossings, two lines based on S97, but static with respect to $p_{SW}$, are fit such that they are either inclusive (dash-dotted lines) or exclusive (dash-dot-dotted lines) of all boundary crossings. These lines are then padded by $20\%$ and used to set upper and lower limits on the possible locations of boundaries (gray shaded).}
    \label{fig:UpperLowerLimits_Spatial}
\end{figure}

These upper- and lower-bounds on the boundary location ($r \rightarrow \infty$ and $r = 0$ for inbound and outbound crossings, respectively) are clearly not physical, as there is a vanishingly small probability that the boundary surface would lie at either $r=0$ or $r \rightarrow \infty$. We instead employ physically motivated upper- and lower-bounds for the inbound and outbound crossings, respectively based on the available crossing data.
To find a physical upper-bound, values of $\mathbf{c_{S97*}}$ in Equation \mbox{\ref{eqn:S97}} are found such that $r_b > r_{crossing}$ holds for all crossing events; likewise, for the physical lower-bound $v_{S97}$ is found such that $r_b < r_{crossing}$ holds for all events. These upper- and lower- bounds thus contain all associated crossings between them, as demonstrated in Figure {\ref{fig:UpperLowerLimits_Spatial}}.
In these fits, we allow only the $r_0$ parameter to vary. Each fit is made static relative to the external pressure ($r_1=\alpha_1=0$), and the flare parameter is fixed as $\alpha_f=\alpha_0=1.0$ for the upper-bound and $\alpha_f=\alpha_0=0.5$ for the lower-bound. Finally, the upper- and lower-bounds found through this fitting process are padded by $20\%$ to account for potential extreme crossings which have been unobserved. These are treated as ``soft'' boundaries (i.e., boundaries which are able to be violated), and so the arbitrary choice of $20\%$ padding here does not preclude even more extreme boundary configurations from being explored.


We then sample all spacecraft trajectories at a $10$-minute cadence, following \citeA{Joy2002}, and calculate the upper- and lower-bounds for each boundary along each trajectory based on the S97* fits; to maintain the tractability of the problem, the dataset is additionally limited such that the maximum difference between the upper- and lower-bounds is $1000 \ \mathrm{R}_\mathrm{J}$. After making this cut to the data, we are left with ${\sim}500,000$ pairs of upper- and lower-bounds between which the boundaries may lie. This dataset is significantly imbalanced, as the spacecraft spend unequal amounts of time inside and outside of each boundary. To address this imbalance and make the model coefficient estimation computationally tractable we sample only those points with $\pm10$ hours of a boundary crossing, plus an equal number of points sampled randomly from all times outside the $\pm10$ hour range. We finally note that, while we attempt to maximize the spacecraft-derived information about the potential locations of each boundary, the lack of magnetopause crossings with $x_{JSS}\lessapprox-90 \ \mathrm{R}_\mathrm{J}$ and bow shock crossing with $x_{JSS}\lessapprox-600 \ \mathrm{R}_\mathrm{J}$ impose strict limits on the validity of the models tested here. The bow shock models become more poorly constrained beyond $x_{JSS}\lessapprox-200 \ \mathrm{R}_\mathrm{J}$, where the density of bow shock crossing events becomes very low; beyond these limits, there are not enough data to validate either boundary model.

\subsection{Bayesian Model Formulation}\label{sec: Bayesian Fitting Methods}

For each boundary shape model $\mathcal{F}_{model}$, we aim to estimate the probability of different values of the parameter vector $\mathbf{c}_{model}$, given the various data and physical models previously discussed. These data and models are the empirical spacecraft position $\mathbf{r}_{sc} = (r_{sc}, \theta_{sc}, \phi_{sc})$ (Section \ref{sec: Spacecraft Measurements}), the estimated skew-normal distribution of solar wind pressures $p_{SW}$ (Section \ref{sec: Solar Wind Model}), and the uniform distribution of possible boundary distances $r_b^{obs}$ (Section \ref{sec: Data Post-Processing}). We infer the parameters composing $\mathbf{c}_{model}$ by formulating a Bayesian model which relates the shapes (e.g., Equations \ref{eqn:S97} and \ref{eqn:new}) to the data and naturally allows the incorporation of the uncertaing estimates of $p_{SW}$ and $r_b^{obs}$ as distributions. The probability of $\mathbf{c}_{model}$ given the data can be formulated with Bayes' theorem \cite{Bayes1763} as
\begin{equation}
    P(\mathbf{c}_{model} | r_{b}^{obs}, \mathbf{x}) = \frac{P(r_b^{obs} | \mathbf{c}_{model}, \mathbf{x}) \; P(\mathbf{c}_{model} | \mathbf{x})}{P(r_b^{obs}|\mathbf{x})}
    \label{eqn: Bayes' Theorem}
\end{equation}
where we have introduced the posterior probability $P(\mathbf{c}_{model} | r_b^{obs}, \mathbf{x})$, the likelihood $P(r_b^{obs} | \mathbf{c}_{model}, \mathbf{x})$, the prior probability $P(\mathbf{c}_{model} | \mathbf{x})$, and the marginal probability $P(r_b^{obs} | \mathbf{x})$. The posterior probability, $P(\mathbf{c}_{model} | \mathbf{x}, r_b^{obs})$, is the conditional probability that $\mathbf{c}_{model}$ defines the boundary shape, given the data $\mathbf{x}$ and $r_b^{obs}$. By inferring this probability distribution, the distributions of possible $\mathbf{c}_{model}$ may be estimated for each model.

The likelihood, $P(r_b^{obs} | \mathbf{c}_{model}, \mathbf{x})$, represents the conditional probability of observing the data given a particular $\mathbf{c}_{model}$. The likelihood can thus be explored numerically, in a similar manner to a forward model: for a particular $\mathbf{c}_{model}$, the likelihood of observing $r_c^{obs}$ can be calculated from its similarity to $r_b^{model} = \mathcal{F}_{model}(\mathbf{c_{model}}, \mathbf{x})$, as described by Equations \ref{eqn:S97} and \ref{eqn:new}. For this analysis, the likelihood is characterized as a normal distribution ($\mathcal{N}(\mu, \sigma^2)$) centered on $r_b^{model}$ with standard deviation $\sigma$ as
\begin{equation}
    P(r_b^{obs} | \mathbf{c}_{model}, \mathbf{x}) = \mathcal{N}(r_b^{obs} | r_b^{model}, \sigma^2) = \mathcal{N}(r_b^{obs} | \mathcal{F}(\mathbf{c}_{model}, \mathbf{x}), \sigma^2)
    \label{eqn: likelihood}
\end{equation}
This standard deviation $\sigma$ is analogous to disagreements between the model ($r_{b}^{model}$) and data ($r_{b}^{obs}$); however, large values of $\sigma$ may result from the choice of simplified shape models, which imperfectly capture the true underlying shape of the boundaries. As $\sigma$ is measured radially outward from the planet, but the variation in boundary location is not likely to be symmetric, we model this standard deviation as
\begin{equation}
    \sigma = \sigma_b + \sigma_m r_{b}^{model} = \sigma_b + \sigma_m \mathcal{F}(\mathbf{c}_{model}, \mathbf{x})
\end{equation}
where $\sigma_b$ represents a constant standard deviation and $\sigma_m$ represents a source of error which changes with radial distance. Modeling $\sigma$ in this way allows the standard deviation to increase with increasing radial distance from the planet.

The prior probability, $P(\mathbf{c}_{model}|\mathbf{x})$, describes our expectation of the possible values of $\mathbf{c}_{model}$, given the state vector $\mathbf{x}$. If we additionally assume that $\mathbf{c}_{model}$ is independent of the spacecraft trajectory and the solar wind pressure, then the prior probability may be simplified further to $P(\mathbf{c}_{model})$. This simplified prior probability, being independent of the measurement value, thus provides an opportunity to encode our physical knowledge of the system into the model. We adopt physically motivated prior distributions for coefficients in $\mathbf{c}_{model}$ based on previous literature values \cite<e.g>{Joy2002} where available. Generally, normal distributions ($\mathcal{N}(\mu, \sigma^2)$) are adopted as priors for parameters which may be negative and gamma distributions ($\Gamma(\mu, \sigma^2)$ are adopted where the parameters are strictly positive. The prior distributions are all defined with standard deviations large enough to permit sufficient exploration of the parameter space; the same prior distributions may thus be used to determine fits to both the magnetopause and the bow shock. The terms describing these prior distributions are recorded in Table \ref{tab: Priors} in \ref{appendix: Prior Distributions}.

The marginal probability, $P(r_b^{obs} | \mathbf{x})$, represents the probability of observing a crossing given the spacecraft trajectory and solar wind conditions, and is difficult to estimate directly. Fortunately the marginal probability is constant, as the posterior probability density distribution must integrate to $1$. By normalizing the numerator on the right-hand-side of Equation \ref{eqn: Bayes' Theorem} (i.e., the product of the likelihood and the prior probability), the marginal probability may safely be ignored in estimating the posterior probability.

To show the efficacy of this formulation, consider the following example: if a spacecraft is in a location where crossings are never expected to occur (e.g., near the planet at midnight LST), then there will be no combination of model parameters $\mathbf{c}_{model}$ for this state $\mathbf{x}$ that will lead to a crossing $r_{b}^{obs}$. In this limiting case, the likelihood $P(r_b^{obs} | \mathbf{c}_{model}, \mathbf{x})$ is effectively independent of $\mathbf{c}_{model}$ and may be simplified to $P(r_b^{obs} | \mathbf{x})$, which is the marginal probability. The likelihood and marginal thus cancel in Equation \ref{eqn: Bayes' Theorem}, and the posterior reduces to the prior probability. More generally, the estimation of the posterior in this formulation is regularized against trajectories with poor sampling for boundary crossings while allowing the incorporation of such trajectories without issue. 

Practically, we estimate the posterior distribution by implementing the ``No-U-Turn-Sampler'' (NUTS) MCMC sampling method  \cite{Hoffman2014} from the \texttt{PyMC} package \cite{Abril-Pla2023}. The MCMC sampler is run for $2.5\times10^3$ steps after tuning for $2\times10^3$ steps in $4$ chains (separate parallel MCMC samplers); all resulting chains are inspected visually to ensure convergence. This yields a total of $1\times10^4$ samples for each parameter vector $\mathbf{c}_{model}$.

We note one shortcoming of this particular Bayesian model implementation: due to the definition used to describe $r_b^{obs}$ as a uniform distribution, estimates for the coefficients of the J02 boundary model described by Equation \ref{eqn: Bayes' Theorem} cannot be sampled. The J02 model is elliptical, meaning it can close at arbitrarily small downtail distances. This definition was not problematic for the fitting methods used by \citeA{Joy2002}, who fit the boundary curves to boundaries identified in 3D MHD simulations of Jupiter's magnetosphere \cite{Ogino1998}, as these simulations have effectively perfect spatial coverage. As the spacecraft data used here do not have complete spatial coverage, and as $r_b^{obs}$ is represented by a uniform distribution with soft bounds, attempting to estimate the posterior distributions of the J02 coefficients tends to result in small, closed boundaries. These boundaries are unphysical, even if they do not violate the limitations imposed by this Bayesian model described here. Nonetheless, we prefer the data-driven empirical method of characterizing Jupiter's magnetospheric boundaries developed here over a simulation-driven method given the wide variety in magnetospheric shapes, scales, and dynamics predicted by different MHD simulations of the system \cite<e.g.>{Ogino1998, Chane2013, Chane2017, Zhang2018, Feng2023}. With additional spatial coverage from the \textit{Juno} extended mission, as well as the upcoming \textit{Europa Clipper} and JUICE missions, a future implementation of these fitting methods may be able to re-fit the J02 boundary models or otherwise improve the models discussed here.

\section{Results and Discussion}\label{sec: Results}
\subsection{Magnetopause}

\begin{figure}[!th]
    \centering
    \includegraphics[width=0.97\textwidth]{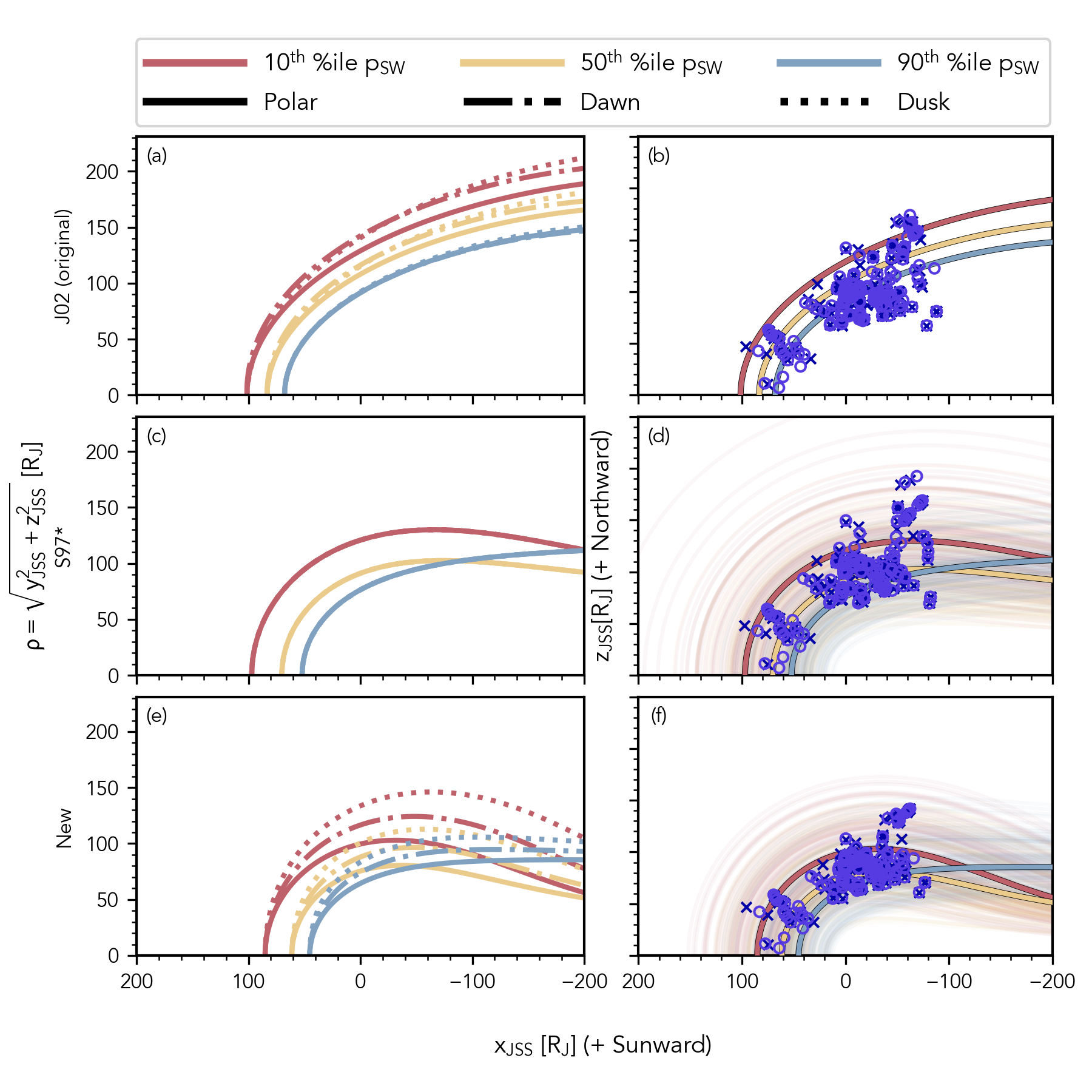}
    \caption{Demonstrations and comparison of each Jovian magnetopause shape model tested. The (a-b) unchanged J02 form, (c-d) S97* form, and (e-f) new form for magnetopause boundary are illustrated as labeled: the first column (a, c, e) shows how the polar regions (solid lines), dawn flank (dot-dashed lines), and dusk flank (dotted lines) vary under solar wind pressures ($p_{SW}$) spanning the $10^{th}$ percentile (red), through the median (yellow) to the $90^{th}$ percentile (blue). For the axisymmetric S97* model, the polar, dawn, and dusk flanks all overlap. The second column (b, d, f) shows representative uncertainties (faint lines) in the boundary shape models for the $\phi=0$ ($y_{JSS} = 0)$ plane, with scaled inbound (purple ``x''s) and outbound (blue ``o''s) crossings as described in the text.}
    \label{fig: Magnetopause Demonstrations}
\end{figure}

For Jupiter's magnetopause, the inferred distributions of the model parameters in $\mathbf{c}_{model}$ are summarized in Table \ref{tab: Magnetopause Parameters}. These values for $\mathbf{c}_{model}$ may be substituted into Equations \mbox{\ref{eqn:S97}} and \mbox{\ref{eqn:new}} to obtain the S97* and new boundary shapes, respectively; these shapes are demonstrated in Figure \ref{fig: Magnetopause Demonstrations} alongside a comparison to the J02 magnetopause shape model. In the second column of Figure \ref{fig: Magnetopause Demonstrations} (i.e., Figures \ref{fig: Magnetopause Demonstrations}b, \ref{fig: Magnetopause Demonstrations}d, and \ref{fig: Magnetopause Demonstrations}f), representative uncertainties $\sigma$ of each magnetopause boundary model are shown for $\phi = 0$, with spacecraft inbound and outbound crossings scaled and overplotted for comparison. The crossing data, which are measured at a range of $\phi$, have been scaled by measuring the ratio of observed-to-modeled crossing distance, or
\begin{equation}
    R = \frac{r_{b}^{obs}}{r_{b}^{model}} = \frac{r_{b}^{obs}}{\mathcal{F}(\mathbf{c}_{model}, \mathbf{x})}
    \label{eqn: R}
\end{equation}
The ratio $R$ is then used to scale the expected distance to the boundary at $\phi=0$ for the same conditions as the crossing data; that is, the scaled crossing distance is given by $\mathcal{F}(\mathbf{c}_{model}, \mathbf{x}') R$ where $\mathbf{x}' = (\theta, \phi=0, p_{SW})$.

\begin{table}[!ht]
    \begin{center}
    \begin{tabular}{crr}
    Param. & S97* & New \\
    \hline
    $r_0$ & $38.0\pm0.0$ & $33.5\pm0.0$ \\
    $r_1$ & $-0.25$ & $-0.25$ \\
    $r_2$ & -- & $12.9\pm0.0$ \\
    $r_3$ & -- & $25.8\pm0.0$ \\
    $a_0$ & $0.28\pm0.00$ & $0.19\pm0.00$ \\
    $a_1$ & $1.08\pm0.00$ & $1.27\pm0.00$ \\
    $\sigma_m$ & $0.15\pm0.01$ & $0.15\pm0.01$ \\
    $\sigma_b$ & $16.2\pm0.8$ & $16.8\pm0.8$ \\
    \end{tabular}
    \caption{Parameters describing the magnetopause shape models}
    \label{tab: Magnetopause Parameters}
    \end{center}
\end{table}

\begin{figure}[!ht]
    \centering
    \includegraphics[width=0.95\textwidth]{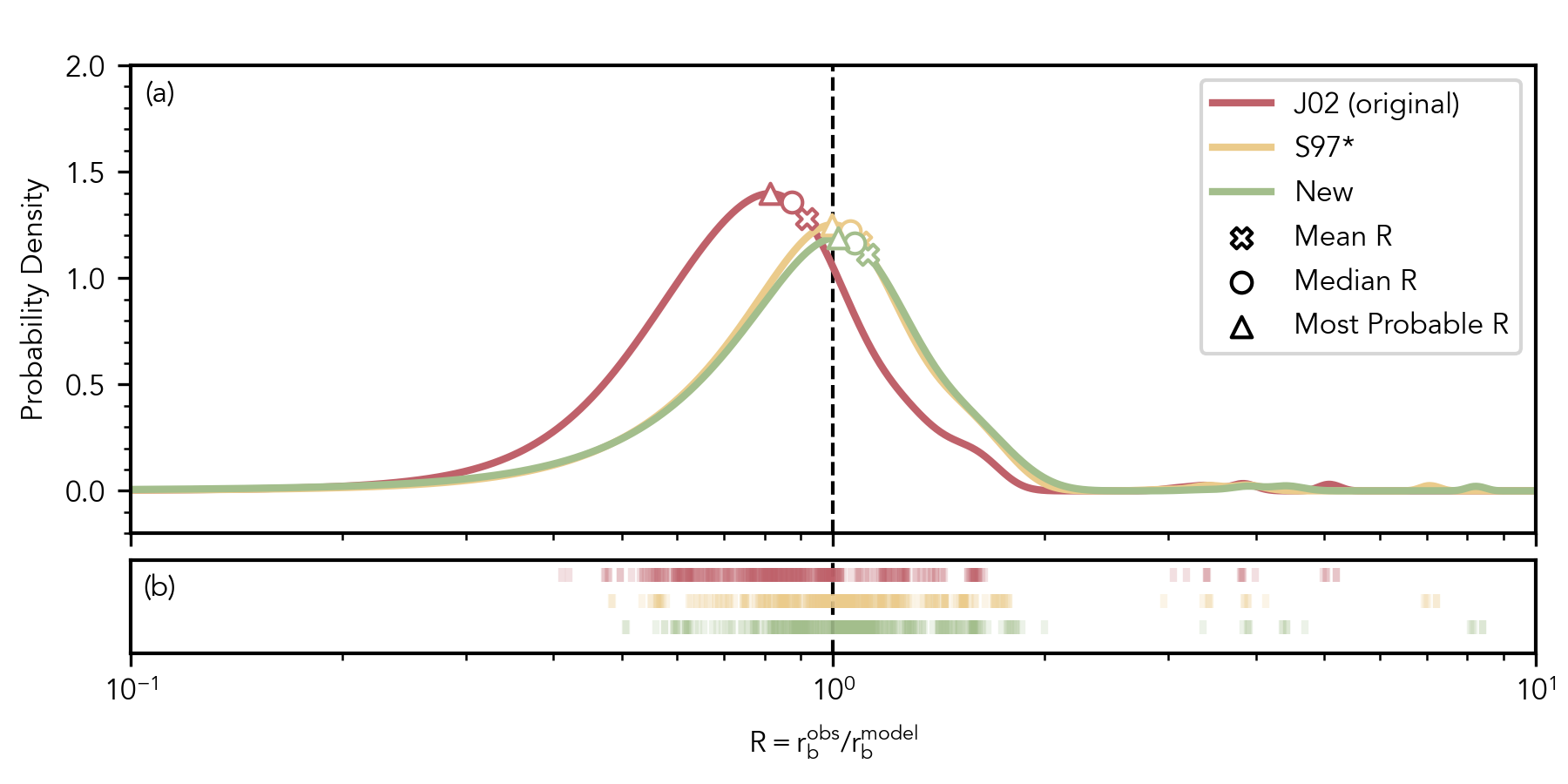}
    \caption{Distributions of the ratio $R = r_{b, data}/r_{b, model}$ for measured-to-modeled magnetopause locations from spacecraft crossing data to the modeled locations, for each model tested here, presented as (a) kernel density estimates and (b) rugplots. $R$-distributions of different models are as labeled (solid lines); for the bimodal models, the $R$-distributions of the inner mode (dashed line), outer mode (dash-dotted line), and weighted sum (solid line) are shown. The mean, median, and most probable (peak) value for each distribution are plotted with symbols of corresponding color. The rugplots instead show each individual $R$ value measured to create each kernel density estimate. A ratio of $R=1$ (black dashed line) denotes perfect data-model agreement; $R>1$ represents model underestimation of the magnetopause location, while $R>1$ represents model overestimation of the magnetopause location.}
    \label{fig: Magnetopause Comparison}
\end{figure}

The performance of the two Bayesian models, the S97* and new magnetopause shapes, may be compared using established statistics, such as leave-one-out (LOO) cross validation \mbox{\cite{Vehtari2017}} or the widely-available information criterion (WAIC) \mbox{\cite{Watanabe2010}}. Both of these statistics effectively characterize the robustness of the model by repeatedly partitioning the data into two sets, using one to estimate the model posterior and the other to test against; here, we report the LOO- and WAIC-derived weights, $w_{LOO}$ and $w_{WAIC}$, respectively, which estimate the relative probability of a Bayesian model being correct. These statistical techniques cannot be applied directly to frequentist (``traditional'' statistical) models, however, making it difficult to compare the J02 model to either the S97* or new models. To get around this limitation, we will use the ratio $R$ as an additional measure of goodness-of-fit for each model, where $R=1$ represents a perfect model prediction of a boundary crossing location, $R>1$ represents an underestimation of the distance to the boundary by the model ($r_{b}^{obs} > r_{b}^{model}$), and $R<1$ represents an overestimation of this distance ($r_{b}^{obs} < r_{b}^{model}$). The distributions of the ratio $R$, calculated for each spacecraft crossing measurement and each magnetopause boundary model, are shown in Figure \ref{fig: Magnetopause Comparison}. The magnetopause models are further compared in Table \ref{tab: Magnetopause Comparison}, which contains a summary of the performance statistics of each model: the mean $\mu_R \pm \sigma_R$, median $M_R$, the most probable value $MPV_R$, and the LOO and WAIC weights, $w_{LOO}$ and $w_{WAIC}$, for the Bayesian models.

\begin{table}[!ht]
    \begin{center}
    \begin{tabular}{lrrr}
    Statistic & J02 & S97 (unimodal) & New (unimodal) \\
    \hline
    $\mu_R\pm\sigma_R$  & $0.92\pm0.08$ & $1.10\pm0.08$ & $ 1.12\pm0.09$  \\
    $M_R$               & $0.88$        & $1.06$        & $1.07$         \\
    $MPV_R$             & $0.82$        & $1.00$        & $1.02$         \\
    \hline
    $w_{LOO}$ & -- & $45.8\%$ & $54.2\%$ \\
    $w_{WAIC}$ & -- & $42.5\%$ & $57.5\%$ \\
    \end{tabular}
    \caption{Magnetopause shape model performance statistics.}
    \label{tab: Magnetopause Comparison}
    \end{center}
\end{table}

These statistical comparisons reveal insights into how the different magnetopause models tested here compare to the crossing data. The J02 model consistently overestimates the scale of the magnetopause, with $R<1$ for approximately ${\sim}70\%$ of all crossings; the ratio $R$ is broadly distributed for the J02 model, indicating a tendency for boundary estimates to be further away from $R=1$. The $R$-distributions from the S97* and new magnetopause models are both centered near $R=1$ and have similar widths; from the $R$ statistics, it is clear that both outperform the J02 model, but not clear whether either of the S97* or new forms outperforms the other. Instead, the Bayesian performance metrics $w_{LOO}$ and $w_{WAIC}$ can be compared: from these, it is clear that the new model is more likely to match the data.

\begin{figure}[!ht]
    \centering
    \includegraphics[width=0.95\textwidth]{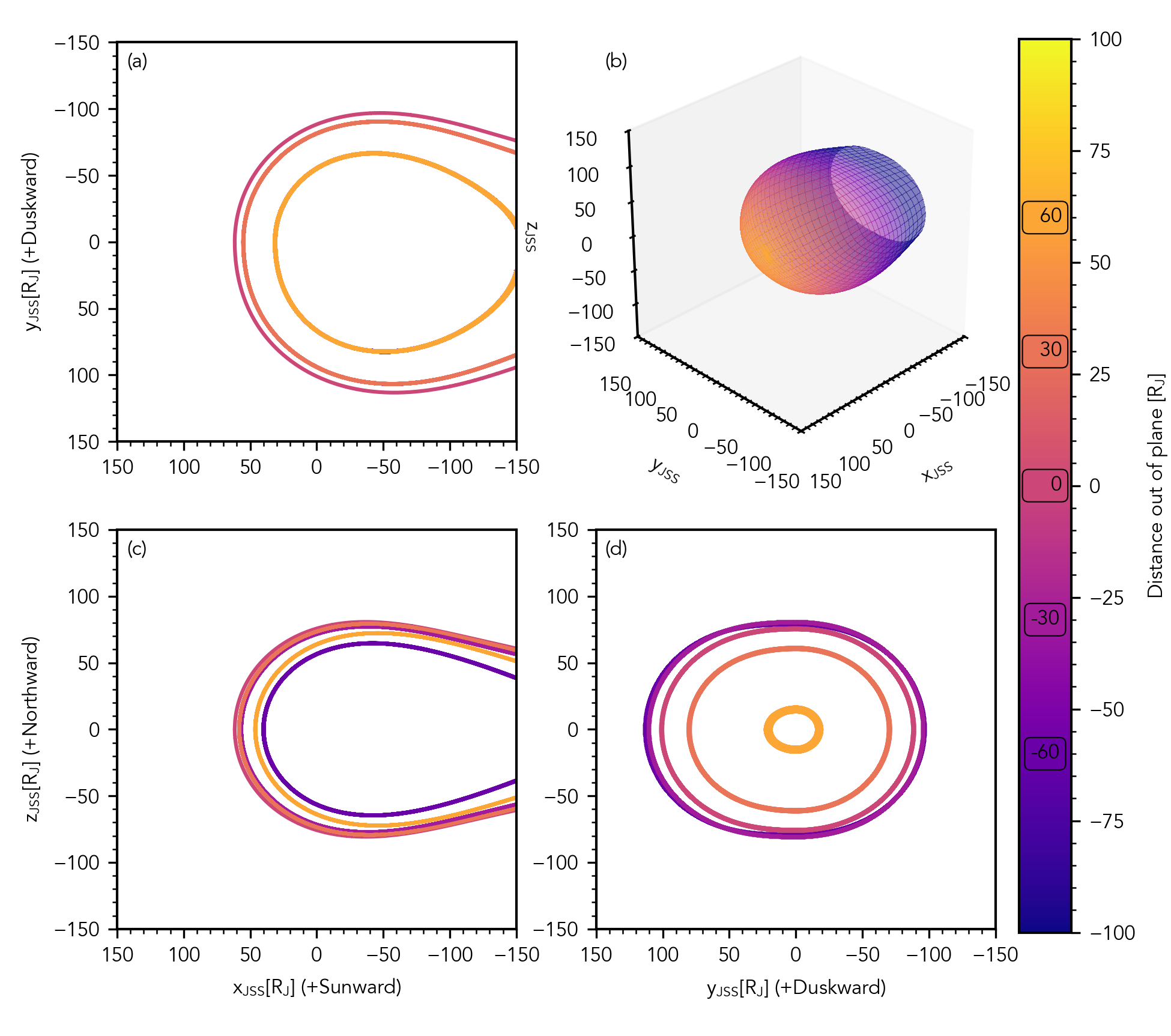}
    \caption{The median best-fit model of the Jovian magnetopause, as described in the text. Panels (a), (c), and (d) show the magnetopause in the $z_{JSS}=const.$, $y_{JSS}=const.$, and $x_{JSS}=const.$ planes, respectively; slices of the magnetopause corresponding to different constant values are differentiated by color, according to the inset labels in the colorbar. Panel (b) shows a 3D perspective of the magnetopause model, colored according to distance from the $x_{JSS}{=}0$ plane and shaded with transparency to reveal the occluded portions of the boundary surface.}
    \label{fig: Best Magnetopause}
\end{figure}

The new magnetopause model, as the best performing model, is shown in more detail in Figure \ref{fig: Best Magnetopause}. This description of the magnetopause differs from the J02 description in several crucial ways: the new form exhibits a higher degree of both polar flattening and of dawn-dusk asymmetry, particularly near the $x_{JSS}=0$ plane and is substantially narrower tailward of the $x_{JSS}=0$ plane, with a more dynamic magnetotail response to changing external pressure. 

Under median external dynamic pressure ($p_{SW}=0.071$ nPa) and in the $x_{JSS}=0$ plane, the J02 model shows polar flattening of $f_{polar}=7\%$ when comparing the polar extent to the dawn flank, while the new model shows $f_{polar}=13\pm47\%$ under the same conditions. The large range of possible values expected for $f_{polar}$ is representative of the highly variable internal pressure components within Jupiter's magnetosphere, which are not explicitly modeled here. The new model is more similar to the $f_{polar}=20\%{\sim}40\%$ range previously estimated \cite<e.g.>{Hill1974, Engle1980, Connerney1981_model, Lepping1981}, and is in particularly good agreement with the expected value of $f_{polar}=22^{+14}_{-22}\%$ recently estimated by \citeA{Ranquist2020}. In the same plane, the J02 model predicts a dawn-to-dusk ratio of $1.01$, indicating a very slight inflation of the dawnside magnetopause relative to dusk; the new model, however, predicts a dawn-to-dusk ratio of $0.93\pm0.41$, instead indicating an inflation of the dusk side relative to dawn. This latter picture qualitatively matches the distribution of spacecraft detections of the magnetosphere summarized in Figure \ref{fig:Residence}c.

Under changing solar wind conditions, the new magnetopause model described here is considerably more dynamic than the J02 model. As $p_{SW}$ increases in magnitude from rarefied conditions (e.g. $p_{SW}=0.02$ nPa, the $10^{th}$ percentile value) through the median value to the conditions of a solar wind compression (e.g. $p_{SW}=0.22$ nPa, the $90^{th}$ percentile value), the subsolar standoff distance of the J02 magnetopause decreases from $r_{SS,10^{th}\%ile}=101 \ \mathrm{R}_\mathrm{J}$ to $r_{SS,50^{th}\%ile}=83 \ \mathrm{R}_\mathrm{J}$ and finally to $r_{SS,90^{th}\%ile}=68 \ \mathrm{R}_\mathrm{J}$. Under the same conditions, the subsolar standoff distance of the new magnetopause boundary described here is consistently more compressed than the J02 model, decreasing from $r_{SS,10^{th}\%ile}=89\pm27 \ \mathrm{R}_\mathrm{J}$ to $r_{SS,50^{th}\%ile}=71\pm24 \ \mathrm{R}_\mathrm{J}$ and finally to $r_{SS,90^{th}\%ile}=51\pm23 \ \mathrm{R}_\mathrm{J}$.

The difference between the J02 shape model and the new model described here is even more evident beyond $x_{JSS}<0$. The J02 model continues to increase in cross section for all values of $p_{SW}$ through the region $0 > x_{JSS} \gtrapprox -300 \ \mathrm{R}_\mathrm{J}$; in the same region, the new model exhibits a shrinking cross section for $p_{SW} < 0.2$ nPa. At incident $p_{SW}$ equal to this critical point ($p_{SW}\approx0.2$ nPa), the new magnetopause model has a constant magnetotail cross section for decreasing $x_{JSS}$; for $p_{SW} > 0.2$ nPa, the new magnetopause model expands with increasing distance downtail. It is important to reiterate, however, that this new model should only be applied for $x_{JSS}\gtrapprox-90 \ \mathrm{R}_\mathrm{J}$, due to the lack of magnetopause crossing constraints tailward of this distance; the asymptotically narrowing tail of the new model is thus unlikely to be a permanent, physical feature of the Jovian magnetopause. 

Nonetheless, just inward of $x_{JSS} = -90 \ \mathrm{R}_\mathrm{J}$, there is a dynamic transition in the new magnetotail model from decreasing cross-section at low $p_{SW}$ to increasing cross-section at high $p_{SW}$. This change in the shape of the magnetopause may be capturing the statistical drop in internal magnetotail pressure associated with magnetic reconnection, either from the Dungey \mbox{\cite{Dungey1961}} or Vasyliunas cycles \mbox{\cite{Vasyliunas1983}}. Signatures of magnetic reconnection have been observed at ${\sim}90 \ \mathrm{R}_\mathrm{J}$ down Jupiter's magnetotail \mbox{\cite<e.g.>{Woch2002, Vogt2010, Vogt2019}}, and the effects of such reconnection on the flow of plasma around Jupiter's nightside magnetosphere have been explored theoretically \mbox{\cite{Cowley2003_polar, Jackman2014}}. In the Vasyliunas cycle in particular, the magnetotail is expected to be extended in the anti-planet direction and plasma-dense before reconnection, which would contribute additional magnetic and thermal pressures near $x_{JSS} = -90 \ \mathrm{R}_\mathrm{J}$; after reconnection, the plasma would be lost downtail as a plasmoid is ejected and the planetary magnetic field would dipolarize, reducing both the magnetic and thermal pressures in the magnetotail. If the rate at which this cycle occurs were anticorrelated with the solar wind dynamic pressure---as might be the case if higher pressure, faster solar wind flows extend the magnetotail in the antisunward direction--- then we might expect the magnetopause to have a shape similar to that described by the new model presented here.

Finally, we note that the magnetopause described by this new model is not explicitly bimodal: all of the parameters described in Table \ref{tab: Magnetopause Parameters} are normally distributed, rather than exhibiting a two-peaked distribution. The distribution of meta-modeled solar wind dynamic pressures $p_{SW}$ from MMESH follow a bimodal distribution themselves, and so the input pressures are unlikely to be at fault. Instead, the lack of a bimodal distribution of magnetopause locations is likely due to the significant standard deviation predicted with the Bayesian model. The $1\sigma$ standard deviation on the new magnetopause model is ${\sim}25 \ \mathrm{R}_\mathrm{J}$ at the subsolar point during median solar wind pressure conditions; while this is comparable to the spread seen in recent MHD models of the Jovian magnetosphere \mbox{\cite{Feng2023}}, such a wide distribution makes recovering a two-peaked density structure difficult. It is possible to force the estimated magnetopause parameters to be bimodal by changing the Bayesian likelihood in Equation \ref{eqn: likelihood} from a normal distribution to a sum of normal distributions; however, parameters estimated after implementing these changes result in models which uniformly agree with the data significantly less than in the single-mode case, and so are not shown here.

\subsection{Bow Shock}

\begin{figure}[!ht]
    \centering
    \includegraphics[width=0.9\textwidth]{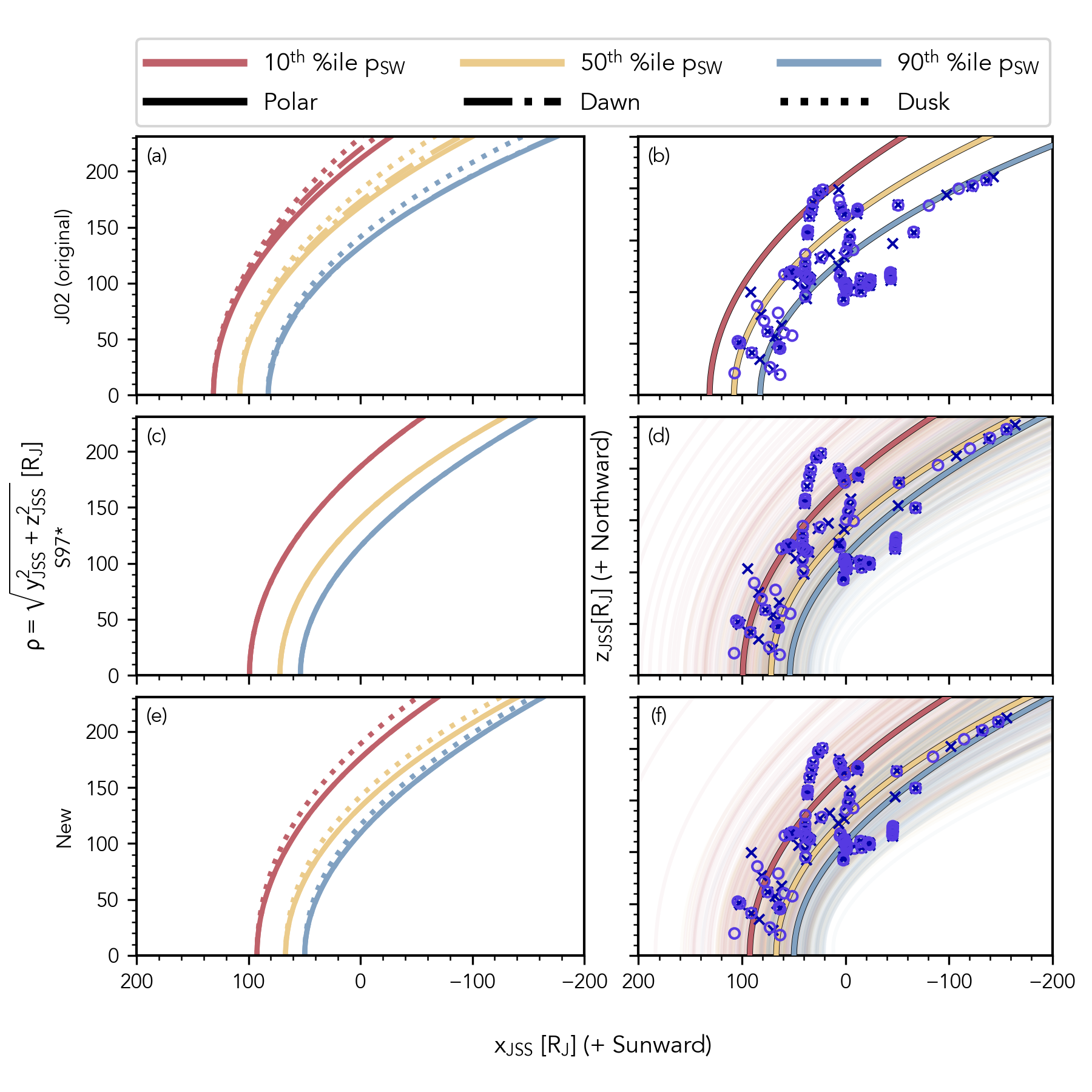}
    \caption{Summary views of each bow shock shape model tested, in the same form as Figure \ref{fig: Magnetopause Demonstrations}: (a-b) the unchanged J02 form, (c-d) the S97* form, and (e-f) the new form.}
    \label{fig: Bow Shock Demonstrations}
\end{figure}

\begin{figure}[!ht]
    \centering
    \includegraphics[width=1.0\textwidth]{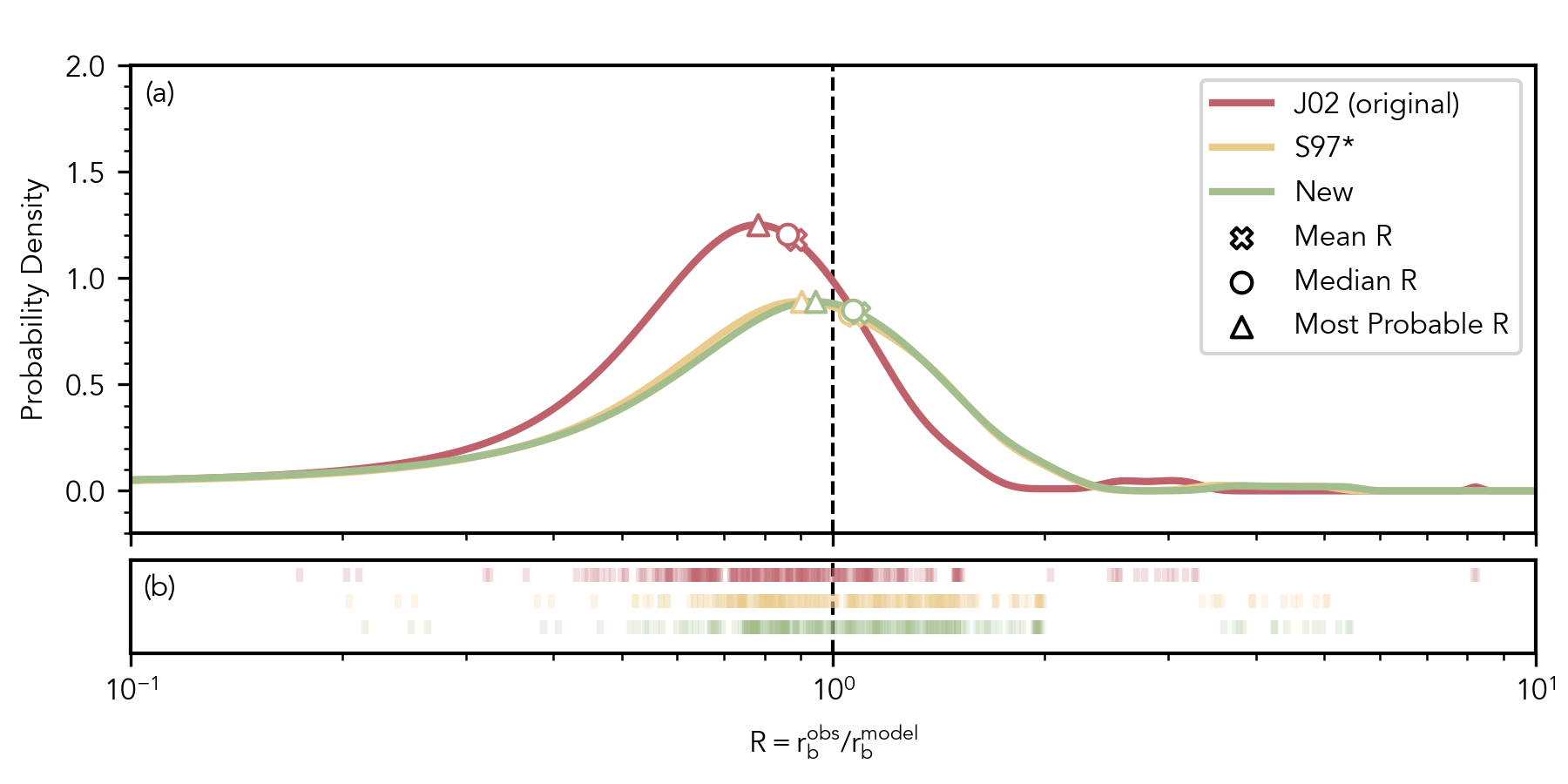}
    \caption{Distributions of the ratio $R = r_{b, data}/r_{b, model}$ in the same format as in Figure \ref{fig: Magnetopause Comparison}, but here for bow shock shape models. As before, a ratio of $R=1$ denotes perfect data-model agreement; $R>1$ represents model underestimation of the magnetopause location, while $R<1$ represents model overestimation of the magnetopause location.}
    \label{fig: Bow Shock Comparison}
\end{figure}

As with the magnetopause shape models, the distributions of probable boundary parameters $\mathbf{c}_{model}$ for the Jovian bow shock are described in Table \ref{tab: Bow Shock Parameters}, with the bow shock shapes corresponding to these parameters demonstrated in Figure \ref{fig: Bow Shock Demonstrations}. The same statistics are employed to measure model performance, including the ratio $R$, as described in Equation \ref{eqn: R}; the distributions of $R$ for each model are shown in Figure \ref{fig: Bow Shock Comparison} and the performance statistics are recorded in Table \ref{tab: Bow Shock Comparison}.

\begin{table}[!htb]
    \begin{center}
    \begin{tabular}{crr}
    Param. & S97* & New \\
    \hline
    $r_0$       & $39.0\pm0.0$  & $36.4\pm0.0$     \\
    $r_1$       & $-0.25$       & $-0.25$          \\
    $r_2$       & --            & $0.0\pm0.0$    \\
    $r_3$       & --            & $9.9\pm0.0$  \\
    $a_0$       & $0.88\pm0.00$ & $0.89\pm0.00$     \\
    $a_1$       & $0.85\pm0.00$ & $0.88\pm0.00$     \\
    $\sigma_m$  & $0.21\pm0.01$ & $0.20\pm0.01$  \\
    $\sigma_b$  & $12.4\pm0.8$. & $13.3\pm0.8$  \\
    \end{tabular}
    \caption{Parameters describing the bow shock shape models.}
    \label{tab: Bow Shock Parameters}
    \end{center}
\end{table}

These statistics reveal a very similar overall picture as was previously described for the magnetopause shape models, despite the apparent differences in the shapes of the $R$-distributions themselves. Again, the J02 model consistently overestimates the scale of the bow shock, with $R<1$ when compared to ${\sim}70\%$ of all measured bow shock crossings. Compared to the $R$-distributions calculated for the magnetopause crossings, those for the bow shock crossings are universally more broadly distributed; the highest probability densities obtained by any model are ${\sim}2.5$ compared to the densities of ${\sim}4$ seen in the magnetopause model comparisons. Among the bow shock models tested here, the J02 model achieves the highest absolute density, and thus is the most narrowly distributed, in stark contrast to what was described for the magnetopause models. As with the bimodal magnetopause models, we see the inner modes very nearly match the corresponding unimodal models, with the outer modes severely overestimating the mean location of the bow shock. This is consistent with the conclusion drawn for the bimodal magnetopause models: the boundaries are not well described by a bimodal in which the mode is determined truly randomly. Interestingly, both unimodal models tested here show signs of an underlying bimodal distribution in the bow shock surface, evidenced by the dual peaks in both $R$-distributions. This apparent bimodal distribution is more clearly defined than any identified in the $R$-distributions of the magnetopause boundary models, despite evidence that the magnetopause should be more clearly bimodal than the bow shock \cite{Joy2002, Collier2020}.

\begin{table}[!htb]
    \begin{center}
    \begin{tabular}{lrrr}
    Statistic & J02 & S97* & New \\
    \hline
    $\mu_R\pm\sigma_R$  & $0.88\pm0.09$ & $1.07\pm0.12$ & $1.09\pm0.12$    \\
    $M_R$               & $0.86$        & $1.40$        & $1.42$           \\
    $MPV_R$             & $0.78$        & $0.90$        & $0.94$           \\
    \hline
    $w_{LOO}$           & -- & $48.6\%$ & $51.4\%$ \\
    $w_{WAIC}$          & -- & $46.9\%$ & $53.1\%$ \\
    \end{tabular}
    \caption{Bow shock shape model performance statistics.}
    \label{tab: Bow Shock Comparison}
    \end{center}
\end{table}

\begin{figure}[!ht]
    \centering
    \includegraphics[width=1.0\textwidth]{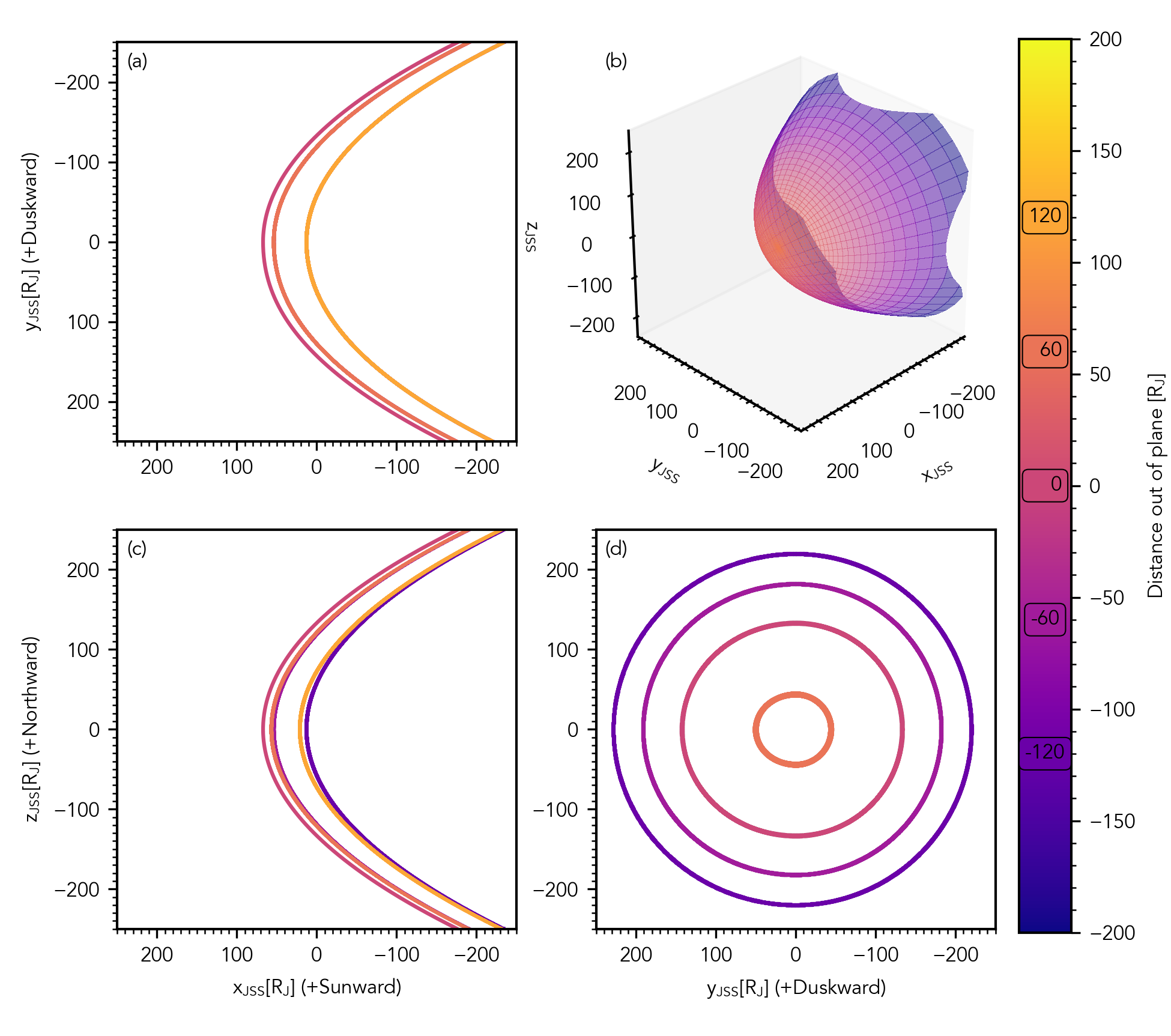}
    \caption{The median best-fit model of the Jovian bow shock, as described in the text, in the same format as in Figure \ref{fig: Best Magnetopause}. Panels (a), (c), and (d) show the bow shock in the $z_{JSS}=const.$, $y_{JSS}=const.$, and $x_{JSS}=const.$ planes, respectively, while panel (b) shows a 3D perspective of the bow shock model, colored according to distance from the $x_{JSS}{=}0$ plane and shaded with transparency to reveal the occluded portions of the boundary surface.}
    \label{fig: Best Bow Shock}
\end{figure}

Taking the Bayesian perfomance statistics into account, the best scoring model in Table \ref{tab: Bow Shock Comparison} is the unimodal version of the new model described here. Compared to the J02 model, this description of the bow shock, shown in detail in Figure \ref{fig: Best Bow Shock}, exhibits a slightly increasing flare angle with increasing solar wind dynamic pressure, growing more blunt as the pressure increases. Within the limits over which the model is valid ($x_{JSS}\gtrapprox-600 \ \mathrm{R}_\mathrm{J}$), however, the magnitude of this change is much smaller than the uncertainties on the model, which are themselves of similar scale to the uncertainties of the most proabable new magnetopause model. This new bow shock model form thus has an effectively fixed flaring angle under changing dynamic pressure $p_{SW}$.

The new bow shock model predicts consistently smaller subsolar standoff distances than the J02 model. As the solar wind dynamic pressure increases from the $10^{th}$ percentile value, through median conditions, to the $90^{th}$ percentile value, the J02 model predicts the subsolar standoff location of the bow shock to evolve from $r_{SS,10^{th}\%ile}=132 \ \mathrm{R}_\mathrm{J}$ to $r_{SS,50^{th}\%ile}=108 \ \mathrm{R}_\mathrm{J}$ and finally to $r_{SS,90^{th}\%ile}=82 \ \mathrm{R}_\mathrm{J}$; the mean ratio of the bow shock standoff distance to that of the magnetopause is thus $R_{SS,BS}/R_{SS,MP}{\approx}1.27$. The unimodal bow shock form described here decreases under the same conditions from $r_{SS,10^{th}\%ile}=100\pm28 \ \mathrm{R}_\mathrm{J}$ through $r_{SS,50^{th}\%ile}=75\pm25 \ \mathrm{R}_\mathrm{J}$ at median pressure conditions before reaching $r_{SS,90^{th}\%ile}=51 \pm 20 \ \mathrm{R}_\mathrm{J}$ at highly compressed conditions. The mean ratio of the bow shock standoff distance to that of the magnetopause for this new model is $R_{SS,BS}/R_{SS,MP}{\approx}1.06\pm0.30$, similar to that of the J02 model and consistent with the results of \citeA{Slavin1985}. It is worth noting that, for very high dynamic pressure conditions $p_{SW} \gtrsim 0.2$ nPa, the new models of the bow shock and magnetopause approach the same subsolar standoff distance. It is unlikely that this situation is physical: in reality, the shocked solar wind downstream of the bow shock will pile up between the bow shock and magnetopause, resulting in some finite buffer between the two boundaries, before flowing around the magnetopause boundary. The uncertainties in the subsolar standoff distance describe this physical situation and allow for the finite distance between the magnetopause and bow shock to be recovered, additionally illustrating the necessity of appropriately considering the uncertainties in any model of planetary magnetospheric boundaries.

In terms of bow shock polar flattening, the J02 model exhibits a flattening of $f_{polar} = 3\%$ under median external dynamic pressure; the new bow shock model shows a comparable flattening of $f_{polar} = 2^{+25}_{-52}\%$. Considering the uncertainties in $f_{polar}$ for the new model, the dawn-to-north polar flattening is very nearly $0\%$, consistent with the nonexistent dawn flank inflation found for the bow shock, as recorded in Table \ref{tab: Bow Shock Parameters}. Due to the uncertainties in the new bow shock model, however, a non-zero dawn-north polar flattening ratio, and thus an inflated dawn bow shock surface, cannot be ruled out entirely. Under the same external conditions and in the $x_{JSS}=0$ plane, the J02 model and new model once again predict opposite directions for the dawn-dusk asymmetry: the J02 model predicts a dawn-to-dusk ratio of $0.95$, while the new model predicts the same ratio to be $1.1\pm0.44$. Interestingly, the directions of these asymmetries are opposite to those predicted by each form for the magnetopause: the J02 magnetopause is inflated at dawn with the bow shock inflated at dusk, while the new model shows that the magnetopause is inflated at dusk with the bow shock inflated at dawn. Once again, however, it must be noted that the large uncertainties in the new model-- representative of the large-scale internal dynamics at Jupiter which modulate the magnetosphere-- mean that all of these dawn-dusk asymmetries are highly variable. At times, either the dawn or dusk flank of either the magnetopause or bow shock may be particularly inflated relative to the average behavior.

\section{Conclusions}
We have combined in-situ detections of different regions around Jupiter-- the magnetosphere, the magnetosheath, and the un-shocked solar wind-- from the \textit{Ulysses}, \textit{Galileo}, \textit{Cassini}, and \textit{Juno} spacecraft with statistical estimates of the contemporaneous, time-varying solar wind conditions from MMESH to test new descriptions of Jupiter's magnetopause and bow shock. In order to effectively leverage the large quantity of spacecraft data, much of which is taken far from the magnetospheric boundaries themselves, and explore the full parameter space for each of these descriptions, we have cast the problem as one of Bayesian statistics and employed an MCMC sampling routine. Two functional forms-- the axisymmetric terrestrial form derived by \citeA{Shue1997}, and a novel asymmetric form based on this-- have been fit to the available data using these methods, and compared to the most recent models describing Jupiter's magnetospheric boundaries from \citeA{Joy2002}. 

The novel, asymmetric boundary shape model developed here-- described throughout the text as the `new' model-- provides the best fit to both Jupiter's magnetopause and bow shock. These models are based on Equation \ref{eqn:new} and are defined by the parameters defined in Tables \ref{tab: Magnetopause Parameters} and \ref{tab: Bow Shock Parameters} for the magnetopause and bow shock, respectively; that is, the magnetopause has the form
\begin{equation}
    \begin{split}
        r_{b, MP} &= \mathcal{F}_{new}(\mathbf{c}_{new,MP}, \mathbf{x}) = r_{SS} \left(\frac{2}{1 + \textrm{cos}(\theta)}\right)^{\alpha_f} + r'_{b}\\
        r_{SS} &= 33.5 * p_{SW}^{-0.25} \\
        \alpha_f &= 0.19 + 1.27 * p_{SW} \\
        r'_{b} &= 
        \begin{cases}
        \sin^2(\theta/2) \left(12.9*\sin^2(\phi)\right) \ p_{SW}^{-0.25},& \text{if } 0 < \phi \leq \pi \\
        \sin^2(\theta/2) \left(25.8*\sin^2(\phi)\right) \ p_{SW}^{-0.25},& \text{if } \pi < \phi \leq 2\pi \\
        \end{cases}
    \end{split}
\end{equation}
with estimated standard deviation $\sigma_{MP} = 16.8 + 0.15 * r_{b,MP}$, while the bow shock has form
\begin{equation}
    \begin{split}
        r_{b, BS} &= \mathcal{F}_{new}(\mathbf{c}_{new,BS}, \mathbf{x}) = r_{SS} \left(\frac{2}{1 + \textrm{cos}(\theta)}\right)^{\alpha_f} + r'_{b}\\
        r_{SS} &= 36.4 * p_{SW}^{-0.25} \\
        \alpha_f &= 0.89 + 0.88 * p_{SW} \\
        r'_{b} &= 
        \begin{cases}
        0,& \text{if } 0 < \phi \leq \pi \\
        \sin^2(\theta/2) \left(9.9*\sin^2(\phi)\right) \ p_{SW}^{-0.25},& \text{if } \pi < \phi \leq 2\pi \\
        \end{cases}
    \end{split}
\end{equation}
with estimated standard deviation $\sigma_{BS} = 13.3 + 0.20 * r_{b, BS}$.
The methods used to derive these boundary forms, and methods which can be used to locate and plot the boundary forms, are publicly available via \url{github.com/mjrutala/JovianBoundaries} \cite{Software_JovianBoundaries}; we encourage the community to use these methods to ensure that uncertainties in the boundary models are accounted for properly.

In general, the new boundary models suggest that Jupiter's magnetospheric boundary surfaces lie much closer to the planet than previous models indicate \cite<e.g>{Ness1979a, Lepping1980, Huddleston1998a, Slavin1985, Joy2002}. Under typical solar wind pressures, the subsolar standoff distance of the new magnetopause is just $r_{SS, MP}=71\pm24 \ \mathrm{R}_\mathrm{J}$, while the same for the bow shock is $r_{SS,BS}=75\pm25 \ \mathrm{R}_\mathrm{J}$. The polar flattening of the magnetopause under the same conditions is $f_{polar}=13\pm42\%$--larger than that predicted by the model of \citeA{Joy2002}-- while the polar flattening of the bow shock is $f_{polar} = 2^{+25}_{-52}\%$--comparable to that predicted by the model of \citeA{Joy2002}. The large uncertainties on these values reflect the high degree of variation in pressure internal to the Jovian magnetosphere; future models could include estimates for the internal magnetic, dynamic, and thermal pressures of Jupiter's magnetosphere in order to reduce these uncertainties. The new model of Jupiter's magnetopause is also substantially more streamlined than the magnetopause model of \citeA{Joy2002}, which may potentially relate to interesting magnetotail dynamics at Jupiter, including reconnection and the associated drops in magnetospheric pressure.

Both boundary models described here have limitations, based primarily on data availability: the magnetopause model is valid for $x_{JSS} > -90 \ \mathrm{R}_\mathrm{J}$ and the bow shock model is valid for $x_{JSS} > -600 \ \mathrm{R}_\mathrm{J}$. Future work to improve these models might relax these limitations by incorporating crossing data from \textit{Voyagers} 1 and 2, which encountered both boundaries further downtail as they left the Jovian system. Additional data from \textit{Pioneer} 10 and 11 and the \textit{Juno} extended mission, or \textit{Europa Clipper} and \textit{JUICE} once these missions reach Jupiter, would further refine these models, as well, provided calibrated solar wind estimates are also available. Finally, solar wind information beyond $p_{SW}$ might be included in an updated model; for instance, the interplanetary magnetic field vector and the Alfvén Mach number ($M_A$) have been shown to modulate the shape and scale of magnetospheric boundaries across the solar system\cite<e.g.>{Peredo1995, Shue1997, Merka2005, Winslow2013}.

\begin{figure}[!ht]
    \centering
    \includegraphics[width=0.9\textwidth]{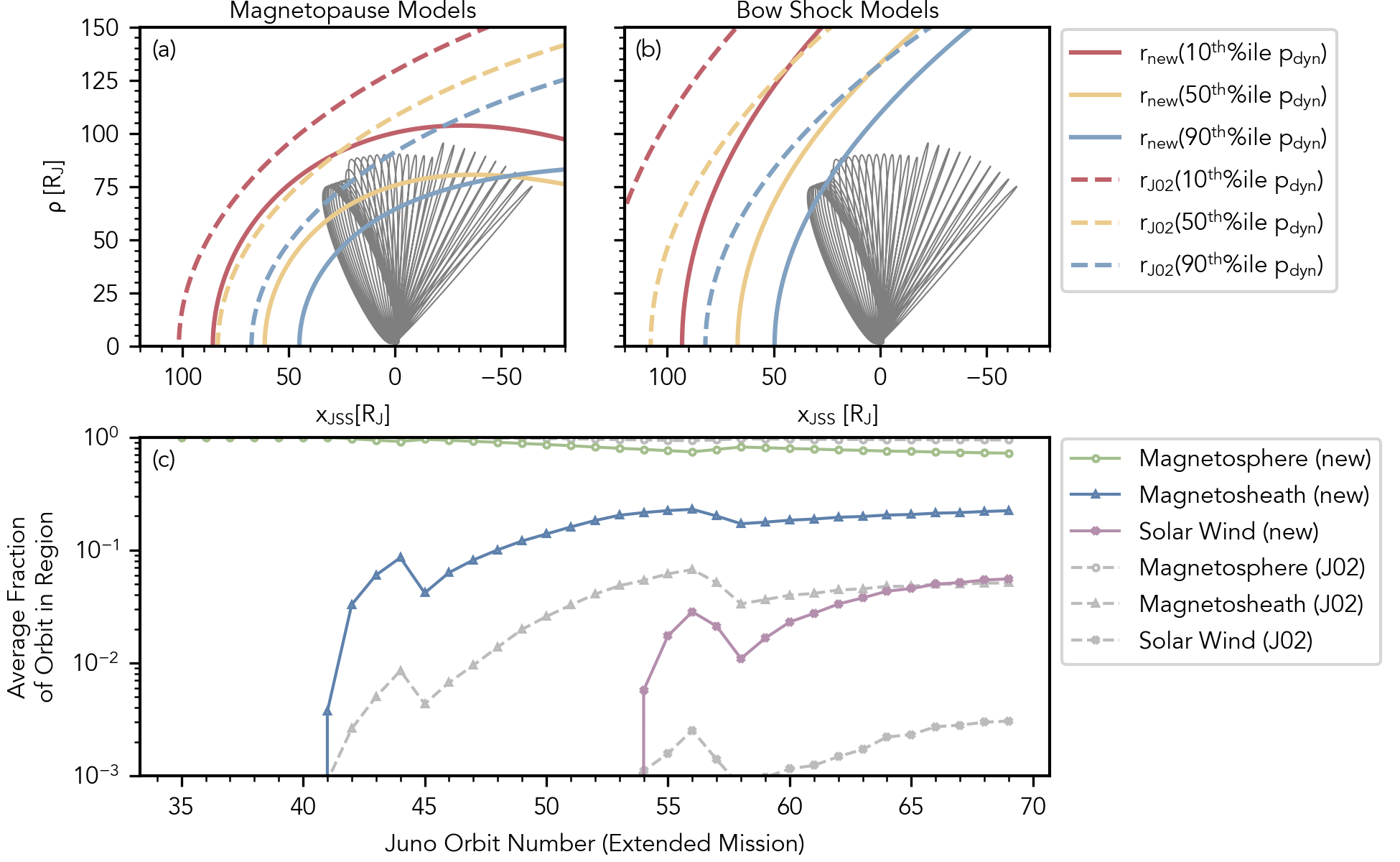}
    \caption{Plots describing how the new boundary models compare to J02 with respect to the \textit{Juno} extended mission, beginning with orbit 35 (21 July 2021). For both the new model (solid lines) and the J02 model (dashed lines), the (a) magnetopause and (b) bow shock forms are shown for the same pressure samples as in Figures \ref{fig: Magnetopause Demonstrations} and \ref{fig: Bow Shock Demonstrations} superimposed on the orbit of the \textit{Juno} spacecraft (gray lines). The new model appears much closer to Jupiter for all pressures, as is validated by looking at (c) the average amount of time spent in each magnetic region (e.g. the magnetosphere, magnetosheath, and solar wind) in each orbit, averaged over the full distribution of solar wind dynamic pressures $p_{SW}$ at Jupiter, as shown in Figure \ref{fig:MMEPerformance}e. The new model predicts that \textit{Juno} will spend less time in the magnetosphere, and more time in the magnetosheath and solar wind, during the later orbits of the extended mission when compared to the J02 models.}
    \label{fig: Juno Prediction}
\end{figure}

These new models of Jupiter's magnetopause and bow shock can be used to predict the approximate portion of each orbit \textit{Juno} will spend in the different magnetic regions surrounding Jupiter: the magnetosphere, magnetosheath, and solar wind. These results are summarized, beginning with the start of the \textit{Juno} extended mission with orbit 35 (2021/07/21-2021/09/02) and continuing through orbit 69 (2025/01/28-2025/03/02), in Figure \ref{fig: Juno Prediction}. Based on the new model, we expect \textit{Juno} to spend a significant fraction of each orbit beyond the magnetopause starting from orbit 42 (2022/05/23-2022/07/05), and beyond the bow shock from orbit 54 (2023/09/07-2023/10/15). By orbit 69 (2025/01/28-2025/03/02), the final orbit for which predicted ephemerides are available, \textit{Juno} is expected to spend roughly ${\sim}72\%$ of its orbit in the magnetosphere, ${\sim}22\%$ of its orbit in the magnetosheath, and ${\sim}6\%$ of its orbit in the solar wind. The J02 model, for comparison, predicts that \textit{Juno} would spend ${\sim}95\%$ of its orbit in the magnetosphere and ${\sim}5\%$ of its orbit in the magnetosheath, with $>1\%$ of its time spent in the solar wind. These predictions can be compared directly to recent magnetospheric boundary crossing measurements made by the \textit{Juno} spacecraft to test the new model presented here; new crossing measurements made by the \textit{Juno} spacecraft can then in turn be used to further refine this new model, particularly in the polar and dusk regions where past missions have not yet explored, but future \textit{Juno} orbits will occur.

\appendix
\section{Updates, Input, and Performance of MMESH}\label{appendix: MMESH}
\subsection{Updates to MMESH}\label{sec: Solar Wind Model: Updates to MMESH}
Several changes have been made to MMESH for the purposes of the analysis discussed here. First, the smoothed F10.7 solar radio flux and solar wind recurrence index are added as parameters to the multiple-linear regression model of arrival time uncertainties. The solar wind recurrence index measures how self-similar the solar wind is over a sidereal solar rotation (characterized by the Carrington rotation period, $\tau_C$), and is calculated as the autocorrelation in the solar wind flow speed $u_{mag}$ over a rolling $\tau_C = 27$ day period \cite{Sargent1985, Zieger2008}. This parameter is calculated from solar wind measurements near the Earth using OMNI data \cite{King2005, Papitashvili2020}, and is further averaged over a 13-Carrington-rotation interval, with half weights on the first and final rotations, to dampen extreme behaviors in the predictor \cite{Hathaway2015}. As the solar wind recurrence index approaches zero, the solar wind is less self-similar and thus more difficult to model, and the meta-modeled arrival time uncertainties should increase. A similar 13-$\tau_C$ rolling mean of the solar F10.7 solar radio flux is also included as a regressor, which is derived from observations at the Dominion Radio Astrophysical Observatory (DRAO) and serves as a proxy for the phase of the solar cycle. 

Second, the distributions of potential values for each meta-modeled solar wind parameter are now parameterized as skew-normal distributions. These distributions are found via a Monte-Carlo (MC) method, wherein the parameter time series is temporally shifted multiple times according to a random sample of the meta-modeled arrival time biases and uncertainties determined by MMESH; the distribution of values of the shifted time series at a particular time then make up these distributions. As the solar wind is structured, a normal distribution of arrival time uncertainties does not translate to a normal distribution of potential values: for instance, when encountering a discontinuity (e.g., a corotating interaction region (CIR)), the rapid change in solar wind characteristics yield asymmetric distributions. As both forward- and backward-shocks are possible in the solar wind, the resulting distributions may be biased to either higher or lower values. The skew-normal thus captures these distributions better than either a normal distribution, due to its symmetry, or a gamma distributions, which only allows bias toward smaller values. Previously, these distributions were not parameterized, but rather characterized by their $16^{th}, 50^{th} (\textrm{median}), \textrm{and } 84^{th}$ percentile values only; parameterizing these distributions allows true random sampling of likely values at each time step.

The MME time series must be similar to the data in both magnitude and variation to successfully model the response of Jupiter's magnetospheric boundaries to solar wind dynamic pressure changes: inconsistency in the magnitude of the dynamic pressure would yield an incorrectly scaled boundary standoff distance; inconsistency in the variance would yield a model which either under- or over-responds to changes in the pressure. To ensure consistency between the ensemble model and data, a correction to the magnitude of the ensemble model is applied to each solar wind parameter at the end of the meta-modeling process. This correction accounts for the MC sampling method used to create distributions for each solar wind parameter, which tends to regress the time-varying model toward the mean, as well as the consistent underestimation of solar wind parameters by physics-based models \cite<e.g.>{Rutala2024_MMESH}. The correction function has the form
\begin{equation}
    f'(t) = m * \bar{f} + s * (f(t) - \bar{f})
\end{equation}
where $f'(t)$ is the corrected value of each parameter as a function of time, $f(t)$ is the original value of each parameter as a function of time, $\bar{f}$ is the mean value of $f(t)$, and $m$ and $s$ are fitting coefficients with forms
\begin{equation}
    m = \frac{\overline{f_{data}}}{\bar{f}}, s = \frac{\sigma_{f_{data}}}{\sigma_{f}}
\end{equation}
where $\overline{f_{data}}$ and $\sigma_{f_{data}}$ are the mean value and standard deviation of the parameter in the data, respectively, and $\sigma_{f}$ is the standard deviation of the parameter in the ensemble model. This correction factor has the effect of scaling the mean value and standard deviation of each solar wind parameter in the ensemble model to that of the data, ensuring similar magnitudes and variation in both. The correction is performed independently for each solar wind parameter, and is iterated over until the mean and standard variation of the ensemble model and data agree to within $1\%$. 

\begin{figure}[!ht]
    \includegraphics[width=\textwidth]{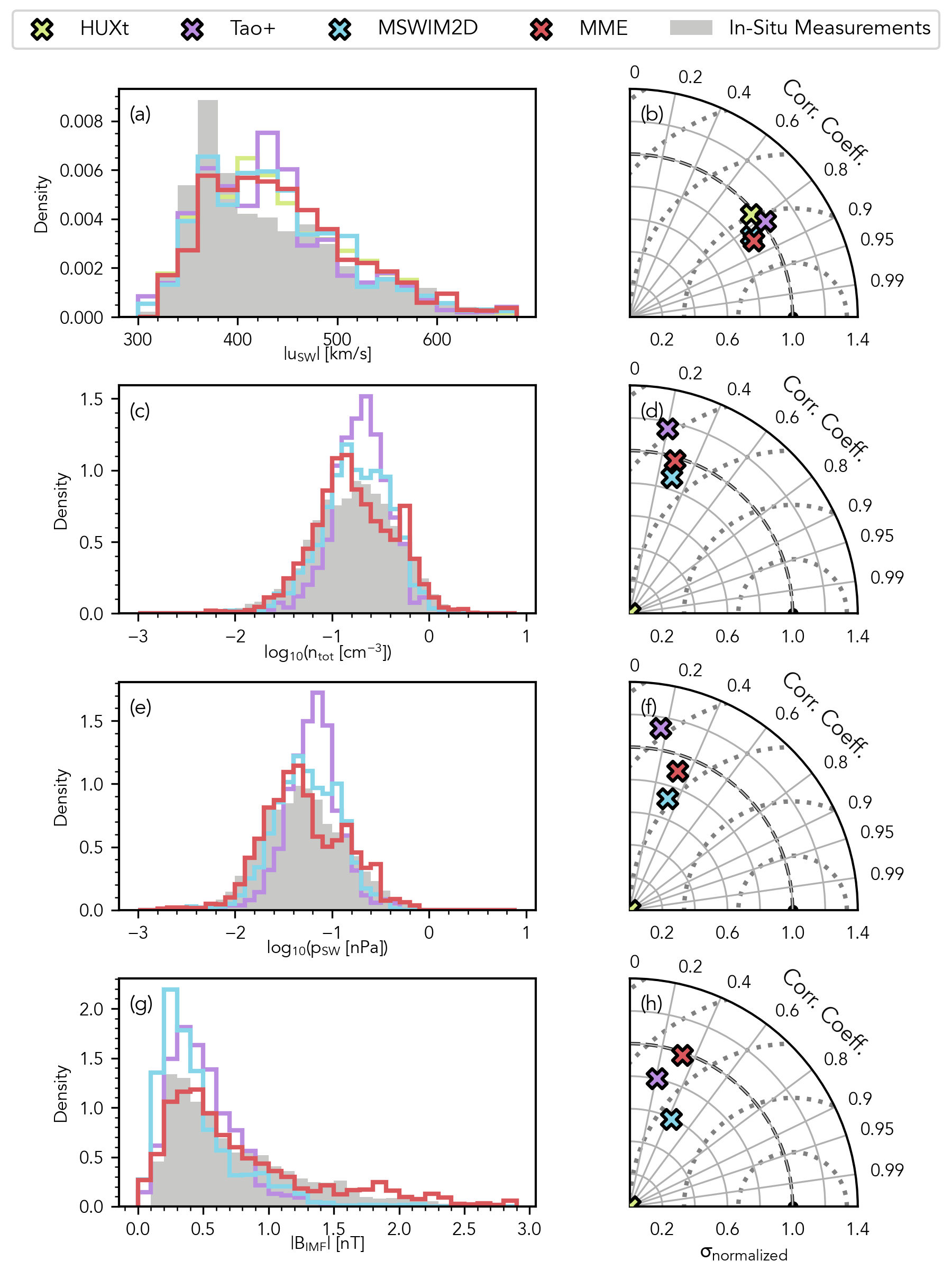}
    \caption{Histograms (a, c, e, g) and Taylor diagrams (b, d, f, h) of each multi-model ensemble (MME) solar wind parameter from MMESH: the solar wind flow speed $u_{mag}$ (a, b); the solar wind density $n_{tot}$ (c, d); the dynamic pressure $p_{SW}$ (e, f); and the strength of the interplanetary magnetic field (IMF) $B_{IMF}$ (g, h). The histograms show the full distribution of solar wind parameter values from the data (gray shading), each input model, and the MME (as labelled), while the Taylor diagrams illustrate the overall standard deviation, correlation coefficient, and root-mean-square-deviation of each model including the MME relative to the data, as described in detail in the text. Taken altogether, these diagrams illustrate how similar the MME solar wind parameters are to the data and illustrate the improvement over any single input model.}
    \label{fig:MMEPerformance}
\end{figure}

\subsection{Input and Performance}\label{sec: Solar Wind Model: Input and Performance}
Following the previously discussed updates to MMESH, a solar wind MME was created using the HUXt \cite{Owens2020, Barnard2022}, MSWIM2D \cite{Keebler2022}, and Tao+ \citeA{Tao2005} solar wind propagation models as input. The temporal uncertainty and performance of each model were measured during epochs where the \textit{Ulysses}, and \textit{Juno} spacecraft were in the solar wind near Jupiter's orbit, spanning distances from the sun $4.90 \textrm{ AU}-5.41 \textrm{ AU}$ and heliolatitudes $^{-}6.1^{\circ}-^{+}6.1^{\circ}$; these same epochs are identical to those used by \citeA{Rutala2024_MMESH}, and are discussed in further detail there. The creation of the MME is otherwise unchanged from that described by \citeA{Rutala2024_MMESH}, save the addition of the solar wind recurrence index and smoothed F10.7 radio flux as predictors in the multiple-linear-regression to the timing uncertainties.

The resulting MME spans the same date ranges as the spacecraft crossings summarized in Table \ref{tab: Crossings Summary}. The performance of the MME is summarized in Figure \ref{fig:MMEPerformance}, where it has been visualized using histograms and Taylor diagrams \cite{Taylor2001}. Briefly, the Taylor diagram uses a polar plot to relate the standard deviation of a model normalized to that of data $\sigma_n$ (radial axis) and the correlation coefficient of the same model to the data $r$ (angular axis). By analogy with the law of cosines, the resulting diagram also illustrates the root-mean-square-deviation ($RMSD$) of a model relative to data (illustrated by the dotted concentric circles in Figure \ref{fig:MMEPerformance}b, d, f, h). On the Taylor diagram, a model which perfectly matches the data would lie at coordinates $(\sigma_n, r) = (1, 1)$, and correspondingly have $RMSD=0$. Most importantly for the magnetospheric boundary models discussed here, the MME dynamic pressure $p_{SW}$, summarized by Figure \ref{fig:MMEPerformance}e-f, matches the distribution of measured dynamic pressures well, and performs better than any individual input model, with $(\sigma_n, r)$ nearer $(1,1)$ than that of either MSWIM2D or Tao+.

\section{Prior Distributions}\label{appendix: Prior Distributions}
The prior distributions for each shape model parameter vector $\mathbf{c}_{model}$ are recorded in Table \ref{tab: Priors}. Any parameters held constant are described by a single number; any parameter described by a distributions is recorded with that distribution and associated parameters. Here, we have made use of the normal distribution ($N(\mu, \sigma)$), gamma distribution ($\Gamma(\mu, \sigma)$), half-normal distribution ($(N_{1/2}(\sigma)$), truncated normal distribution ($N_T(\mu, \sigma, l)$), and Dirichlet distribution ($D(a_{n})$).

\begin{table}[!ht]
    \begin{center}
    \begin{tabular}{crrrr}
    Param. & S97 (unimodal) & New (unimodal) \\
    \hline
    $r_0$       & $G(\mu=40, \sigma=20)$    & $G(\mu=40, \sigma=20)$    \\
    $r_1$       & $-0.25$                   & $-0.25$                   \\
    $r_2$       & --                        & $G(\mu=10, \sigma=10)$    \\
    $r_3$       & --                        & $G(\mu=10, \sigma=10)$    \\
    $a_0$       & $G(\mu=1, \sigma=0.5)$    & $G(\mu=1, \sigma=0.5)$    \\
    $a_1$       & $N(\mu=0, \sigma=1)$.     & $N(\mu=0, \sigma=1)$      \\
    $\sigma_m$  & $N_{1/2}(\sigma=0.01)$    & $N_{1/2}(\sigma=0.01)$    \\
    $\sigma_b$  & $N_{1/2}(\sigma=1.0)$     & $N_{1/2}(\sigma=1.0)$     \\
    \end{tabular}
    \caption{Prior distributions for all shape model parameters.}
    \label{tab: Priors}
    \end{center}
\end{table}

Appropriate prior distributions were determined based on the physical interpretation of each parameters. The normal distribution $N(\mu, \sigma)$ was used for any parameter which may take a negative value, while the gamma distribution $\Gamma(\mu, \sigma)$ was used for most parameters which are strictly positive. The half-normal distribution $(N_{1/2}(\sigma)$, which has maximum likelihood at $0$ and is described only by a standard deviation $\sigma$, was used as a prior on the model uncertainties, $\sigma_m$ and $\sigma_b$, as is standard for estimating uncertainity in an MCMC sampler.

\section*{Open Research}
Spacecraft data in the form of magnetopause crossing event lists were obtained from \citeA{Bame1992} (\textit{Ulysses}), \citeA{Kurth2002} (\textit{Galileo} and \textit{Cassini}), \citeA{Svenes2004} (\textit{Cassini}), \citeA{Hospodarsky2017} (\textit{Juno}), and \citeA{Louis2023} (\textit{Juno}). Bow shock crossing event lists were obtained from \citeA{Bame1992} (\textit{Ulysses}), \citeA{Kurth2002} (\textit{Galileo} and \textit{Cassini}), \citeA{Achilleos2004} (\textit{Cassini}), \citeA{Hospodarsky2017} (\textit{Juno}), and \citeA{Louis2023} (\textit{Juno}). Additional, previously unpublished \textit{Galileo} boundary crossing events were obtained from the PhD thesis of \citeA{Joy2010}; these crossings have been reproduced in Supporting Information (SI) 1. All spacecraft data were supplemented with ephemerides from the NASA Navigation and Ancillary Information Facility (NAIF) SPICE toolkit (https://naif.jpl.nasa.gov/naif/). Component solar wind models used in MMESH were obtained from AMDA \cite{Genot2021} (Tao+, https://amda.irap.omp.eu/), \citeA{Keebler2022} (MSWIM2D, http://csem.engin.umich.edu/MSWIM2D/), or run locally from open-source code \cite{Owens2020, Barnard2022} (https://github.com/University-of-Reading-Space-Science/HUXt).

All code used in this manuscript, including the data processing, Bayesian estimation, and plotting, is published \cite{Software_JovianBoundaries}. The MMESH results used to estimate the solar wind dynamic pressure are similarly available \cite{Dataset_Rutala2025}, as is the MMESH code itself \cite{Software_MMESH_0.2.0}.






\acknowledgments
The work of MJR and CMJ at DIAS is supported by Taighde Éireann - Research Ireland award 18/FRL/6199. The work of CFB at DIAS is supported by the Taighde Éireann - Research Ireland Laureate Consolidator award SOLMEX.


%
%



\bibliography{JovianBoundaries}

%
%
%
%
%

\end{document}